\documentclass[twocolumn]{emulateapj}

\usepackage{comment}
\usepackage{color}
\usepackage{soul}
\usepackage{amssymb,amsmath}

\shorttitle{X-ray Selected Galaxy Groups in Bo\"otes}
\shortauthors{Vajgel et al.}

\begin{document}


\title{X-ray Selected Galaxy Groups in Bo\"otes}


\author{
Bruna Vajgel\altaffilmark{1,2},
Christine Jones\altaffilmark{2},
Paulo A. A. Lopes\altaffilmark{1},
William R. Forman\altaffilmark{2},
Stephen S. Murray\altaffilmark{2,3},
Andrew Goulding\altaffilmark{2},
Felipe Andrade-Santos\altaffilmark{2}
}
\affil{$^1$Observat\'orio do Valongo - Universidade Federal do Rio de Janeiro,
    Rio de Janeiro - RJ, Brazil}
\affil{$^2$Harvard-Smithsonian Center for Astrophysics, 60 Garden
  Street, Cambridge, MA 02138, USA}   
\affil{$^3$Department of Physics and Astronomy, John Hopkins University, Baltimore, MD 21218, USA}



%%%%%%%%%%%%%%%%%%%%%%%%%%%%%%%%%%%%%%%%%%%%%%%%%%%%%%%%%%%%%%%%%%%%%%%%
%%%%%%%%%%%%%%%%%%%%%%%%%%%%%%%%%%%%%%%%%%%%%%%%%%%%%%%%%%%%%%%%%%%%%%%%
%%%%%%%%%%%%%%%%%%%%%%%%%%%%%%%%%%%%%%%%%%%%%%%%%%%%%%%%%%%%%%%%%%%%%%%%
%%%                                                                  %%%
%%%                           ABSTRACT                               %%%
%%%                                                                  %%%
%%%%%%%%%%%%%%%%%%%%%%%%%%%%%%%%%%%%%%%%%%%%%%%%%%%%%%%%%%%%%%%%%%%%%%%%
%%%%%%%%%%%%%%%%%%%%%%%%%%%%%%%%%%%%%%%%%%%%%%%%%%%%%%%%%%%%%%%%%%%%%%%%
%%%%%%%%%%%%%%%%%%%%%%%%%%%%%%%%%%%%%%%%%%%%%%%%%%%%%%%%%%%%%%%%%%%%%%%%

\begin{abstract}
We present the X-ray and optical properties of the galaxy groups selected in the Chandra X-Bo\"otes survey. We used follow-up Chandra observations to better define the group sample and their X-ray properties. Group redshifts were measured from the AGES spectroscopic data. We used photometric data from the NOAO Deep Wide Field Survey (NDWFS) to estimate the group richness ($N_{gals}$) and the optical luminosity ($L_{opt}$). Our final sample comprises 32 systems at \textbf{$z<1.75$}, with 14 below $z = 0.35$. For these 14 systems we estimate velocity dispersions ($\sigma_{gr}$) and perform a virial analysis to obtain the radii ($R_{200}$ and $R_{500}$) and total masses ($M_{200}$ and $M_{500}$) for groups with at least five galaxy members. We use the Chandra X-ray observations to derive the X-ray luminosity ($L_X$). We examine the performance of the group properties $\sigma_{gr}$, $L_{opt}$ and $L_X$, as proxies for the group mass. Understanding how well these observables measure the total mass is important to estimate how precisely the cluster/group mass function is determined. Exploring the scaling relations built with the X-Bo\"otes sample and comparing these with samples from the literature, we find a break in the $L_X$-$M_{500}$ relation at approximately $M_{500} = 5\times10^{13}$ M$_\odot$ (for $M_{500} > 5\times10^{13}$ M$_\odot$, $M_{500} \propto L_X^{0.61\pm0.02}$, while for $M_{500} \leq 5\times10^{13}$ M$_\odot$, $M_{500} \propto L_X^{0.44\pm0.05}$). Thus, the mass-luminosity relation for galaxy groups cannot be described by the same power law as galaxy clusters. A possible explanation for this break is the dynamical friction, tidal interactions and projection effects which reduce the velocity dispersion values of the galaxy groups. By extending the cluster luminosity function to the group regime, we predict the number of groups that new X-ray surveys, particularly eROSITA, will detect. Based on our cluster/group luminosity function estimates, eROSITA will identify $\sim$1800 groups ($L_X = 10^{41}-10^{43}$ ergs s$^{-1}$) within a distance of 200 Mpc. Since groups lie in large scale filaments, this group sample will map the large scale structure of the local universe.
\end{abstract}

\keywords{galaxies: groups: general - surveys - X-rays: galaxies: groups}

%%%%%%%%%%%%%%%%%%%%%%%%%%%%%%%%%%%%%%%%%%%%%%%%%%%%%%%%%%%%%%%%%%%%%%%%
%%%%%%%%%%%%%%%%%%%%%%%%%%%%%%%%%%%%%%%%%%%%%%%%%%%%%%%%%%%%%%%%%%%%%%%%
%%%%%%%%%%%%%%%%%%%%%%%%%%%%%%%%%%%%%%%%%%%%%%%%%%%%%%%%%%%%%%%%%%%%%%%%
%%%                                                                  %%%
%%%                           INTRODUCTION                           %%%
%%%                                                                  %%%
%%%%%%%%%%%%%%%%%%%%%%%%%%%%%%%%%%%%%%%%%%%%%%%%%%%%%%%%%%%%%%%%%%%%%%%%
%%%%%%%%%%%%%%%%%%%%%%%%%%%%%%%%%%%%%%%%%%%%%%%%%%%%%%%%%%%%%%%%%%%%%%%%
%%%%%%%%%%%%%%%%%%%%%%%%%%%%%%%%%%%%%%%%%%%%%%%%%%%%%%%%%%%%%%%%%%%%%%%%

\section{Introduction}
\label{introduction}

\par Galaxy groups have lower masses, lower velocity dispersions, lower luminosities, and smaller extents than galaxy clusters.  However, galaxy groups are not simply scaled down versions of rich clusters (e.g. \citet{2000Mulchaey, 2003Ponman, 2005Voit}).  Due to a group's shallow gravitational potential, feedback processes (e.g. galactic winds and AGN feedback) play important roles in the group's evolution. Feedback processes also can increase systematically the intrinsic scatter and change global properties.  Because the evolution of galaxy groups is not a simple product of gravitational mechanisms, it is complex to reproduce it in simulations and thus necessary to check the results with observations.

\par The matter composition in groups also may be altered by feedback processes.  While, in clusters, the intracluster medium (ICM) is strongly dominated by the hot gas, in groups the mass of the galaxy members can exceed the gas mass \citep{2009Giodini}.  When this occurs, the characteristic properties of the gas, including X-ray luminosity ($L_X$), X-ray temperature ($T_X$) and gas mass fraction ($f_g$), will be lower compared to the dynamical properties, including velocity dispersion ($\sigma$) and total mass ($M_{Tot}$).  A direct consequence of this is the break in the scaling relations of galaxy clusters and galaxy groups \citep{2008Dave, 2009Pope, 2011Mittal}.

\par Scaling relations for clusters and groups including X-ray luminosity vs. mass, X-ray luminosity vs. velocity dispersion and X-ray luminosity vs. temperature, have been investigated extensively \citep{2001Finoguenov, 2002Reiprich, 2006Popesso, 2008Rykoff, 2009bLopes, 2009Vikhlinin, 2010Ettori, 2010Leauthaud}. However the question of whether the relations determined for clusters also hold true for poor clusters and groups still remains unsettled.  

\par Some evidence supports a break in the scaling relations at the low end of the group/cluster mass range, possibly caused by the the strong influence of non-gravitational physics on low-mass groups. In studies of poor systems, \citet{2000Mahdavi} and \citet{2000Xue} found $L_X \propto \sigma^{1.38\pm 0.4}$ and $L_X \propto \sigma^{2.35 \pm 0.21}$, flatter than the theoretical expectation for a self-similar model, $L_X \propto \sigma^4$.  \citet{2000Helsdon} ($L_X \propto \sigma^{2.4 \pm 0.4}$) and \citet{2004Osmond} ($L_X \propto \sigma^{2.31 \pm 0.62}$) also found significantly flatter relations in groups.  For groups, \citet{2001Finoguenov} reported a steeper $M-T$ slope than for clusters.  \citet{2012Maughan} found an observed steepening in the $L_X-T_X$ relation for relaxed systems below 3.5 keV and argued it is caused by central heating that affects the intracluster medium (ICM) to larger radii in lower mass systems.

\par However, other studies do not support a break in the scaling relations between groups and clusters, but instead find consistent results for the whole observed mass range, although often with larger scatter for groups for the $L_X-\sigma$, $L_X-T$, $M-T$, $\sigma-T$ and $M-Y_X$ relations (e.g. \citet{1998Mulchaey}; \citet{2004Osmond}; \citet{2009Sun}; \citet{2011Eckmiller}). 

\par For future surveys like eROSITA \citep{2010Predehl} that will detect up to a hundred thousands clusters, it is essential to use reliable mass-observable proxies to estimate the mass regime of the detected clusters and groups.  Thus our main goal is to determine the scaling relations and Log\texttt(N)-Log\texttt(S) for a complete sample of X-ray selected groups limited in flux and to examine scaling relations below $z \le 0.35$ to determine if there is a break in these relations between clusters\footnote{In our survey we do not expect to observe many clusters with $L_X > 2\times10^{44}$ ergs s$^{-1}$, since their numeric density is equal to 10$^{-6}$ Mpc$^{-3}$ and our Chandra survey observations are only 5 ks and cover 9 square degree.} and groups.

\par This paper is organized as follows: in $\S$2 we present the observations and the sample; in $\S$3 we describe the methods to estimate the group's redshift, velocity dispersion ($\sigma_{gr}$), virial radius ($R_{500}$) and mass ($M_{500}$); in $\S$4  we determine the optical luminosity ($L_{opt}$) and richness ($N_{gals}$); in $\S$5 the  X-ray ($L_X$ and $T_X$) properties are derived; in $\S$6 and $\S$7 we discuss the scaling-relations and Log\texttt(N)-Log\texttt(S) results and in $\S$8 we present the conclusions. The cosmology assumed in this work is $\Omega_m=$0.3, $\Omega_{\Lambda}=$0.7, and $H_0=$ 100 \textit{h} km s$^{-1}$ Mpc$^{-1}$, with \textit{h} set to 0.7.

%%%%%%%%%%%%%%%%%%%%%%%%%%%%%%%%%%%%%%%%%%%%%%%%%%%%%%%%%%%%%%%%%%%%%%%%
%%%%%%%%%%%%%%%%%%%%%%%%%%%%%%%%%%%%%%%%%%%%%%%%%%%%%%%%%%%%%%%%%%%%%%%%
%%%%%%%%%%%%%%%%%%%%%%%%%%%%%%%%%%%%%%%%%%%%%%%%%%%%%%%%%%%%%%%%%%%%%%%%
%%%                                                                  %%%
%%%                            OBSERVATIONS                          %%%
%%%                                                                  %%%
%%%%%%%%%%%%%%%%%%%%%%%%%%%%%%%%%%%%%%%%%%%%%%%%%%%%%%%%%%%%%%%%%%%%%%%%
%%%%%%%%%%%%%%%%%%%%%%%%%%%%%%%%%%%%%%%%%%%%%%%%%%%%%%%%%%%%%%%%%%%%%%%%
%%%%%%%%%%%%%%%%%%%%%%%%%%%%%%%%%%%%%%%%%%%%%%%%%%%%%%%%%%%%%%%%%%%%%%%%

\section{Observations}
\label{Observations}

\begin{figure*}
\begin{center}
\includegraphics[width=154mm]{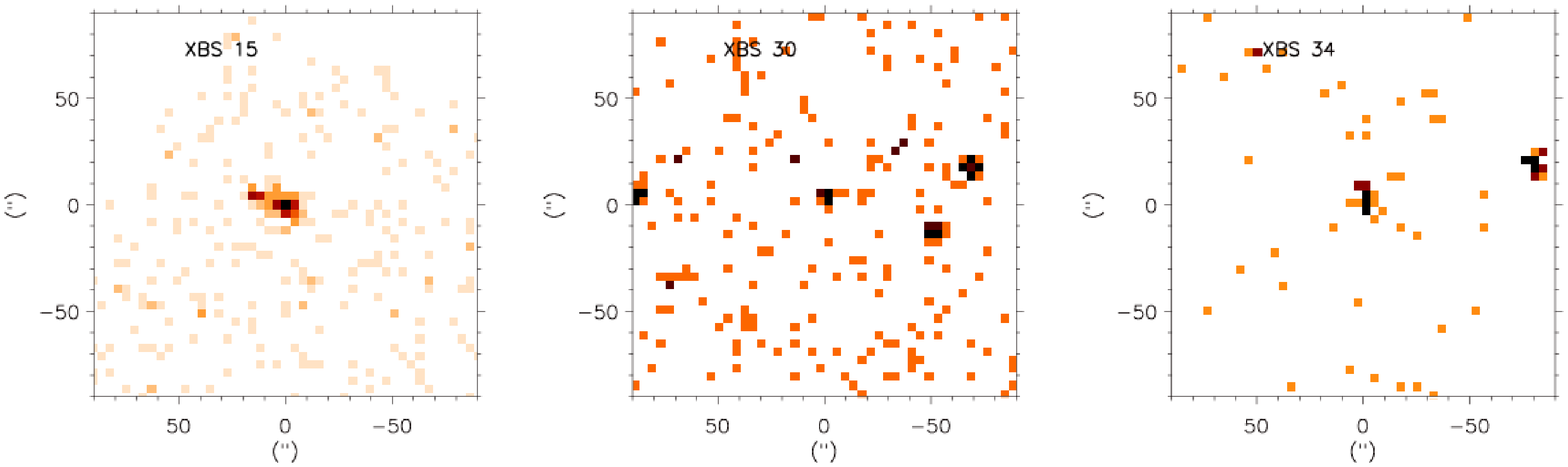}
\end{center}
\caption{Chandra images, in the 0.5-2 keV energy band binned with 4'' resolution, for three sources originally classified by \citet{2005Kenter} as extended, that we found in deeper Chandra observations to be point-like.}
\label{ps_example}
\end{figure*}

\begin{figure*}
\begin{center}
\includegraphics[width=154mm]{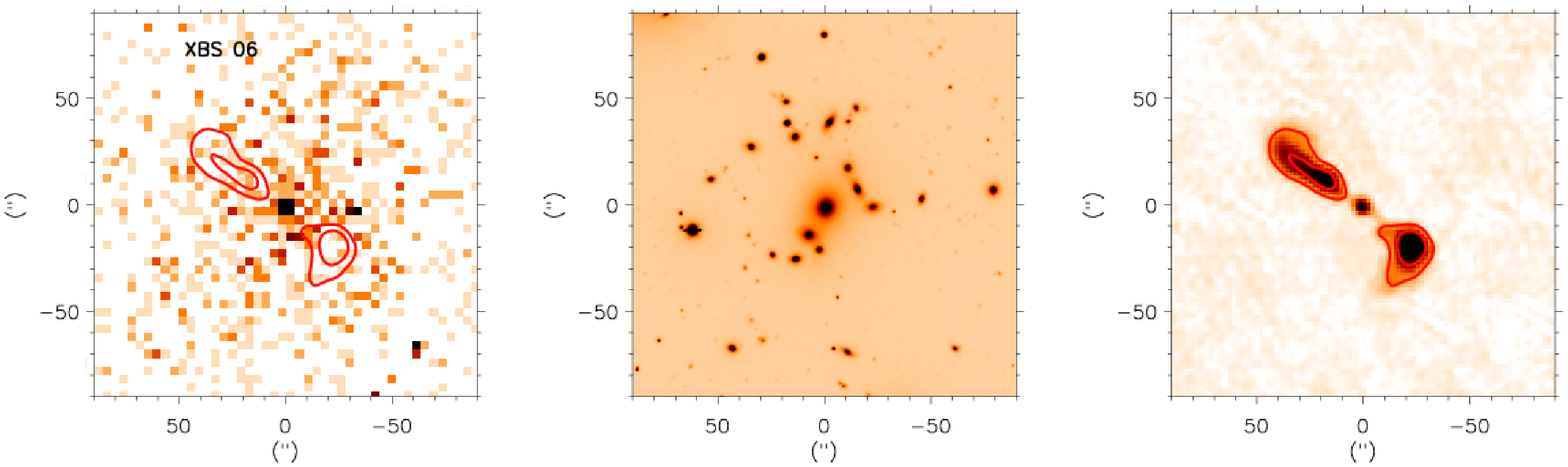}
\end{center}
\caption{The extended X-ray source XBS 06 at $\alpha =$ 14:26:57.9 and $\delta =$ 34:12:01.  This group has 2300 net X-ray counts, $L_{X} = 3.1 \times 10^{43}$ ergs/s in the 0.5-2 keV band and $N_{gals} = 50$.  Left panel shows the ACIS-S 30 ks image (ObsId = 10495) in 0.5-2 keV band with the radio contours of the jets (red lines).  It is possible to see a cavity in the X-ray emission, probably associated with the SW radio lobe.  Middle panel presents  the I-band image from NDWFS, where we can see the BCG at the center of the X-ray emission.  Right panel is the FIRST-NRAO radio image which shows the BCG's radio lobes, which extend $\approx$ 110 kpc and 75 kpc.}
\label{es_example}
\end{figure*}

\par In this section, we describe the three sets of observations covering the Bo\"otes region used in this paper to extract and estimate the properties of the X-ray detected galaxy groups.  These observations include a deep six band optical and infrared photometric survey \citep{1999Jannuzi}, the Chandra X-Bo\"otes survey \citep{2005Murray} and the optical spectroscopic survey of AGNs and galaxies brighter than $I = 20$ \citep{2012Kochanek}.

\subsection{NOAO Deep Wide Field Survey (NDWFS)}
\label{ndwfs}

\par The NOAO Deep Wide Field Survey (NDWFS, \citet{1999Jannuzi}) is a deep optical and IR (BwRIJHK) survey, mapping a total area of 18 deg$^2$ (two regions of 9 deg$^2$ each, one in the Bo\"otes constellation and the other in the Cetus constellation) to faint flux limits (BwRI $\leq$ 26 AB mag; JH = 21 and K = 21.4 at 5$\sigma$ detection limits).  The optical imaging was done with the wide field (36' $\times$ 36') MOSAIC cameras on NOAO's 4 m telescope.  The IR imaging was done with the Ohio State/NOAO Imaging Spectrograph (ONIS) on the KPNO 2.1 m.

\par In this paper, we are interested in the northern Bo\"otes field, which covers 3$^\circ$ $\times$ 3$^\circ$ and is centered at (J2000) RA = 14:32:06 and Dec = +34:16:48.

\subsection{X-Bo\"otes Survey}
\label{xbootes}

\begin{figure*}
\begin{center}
\includegraphics[width=114mm]{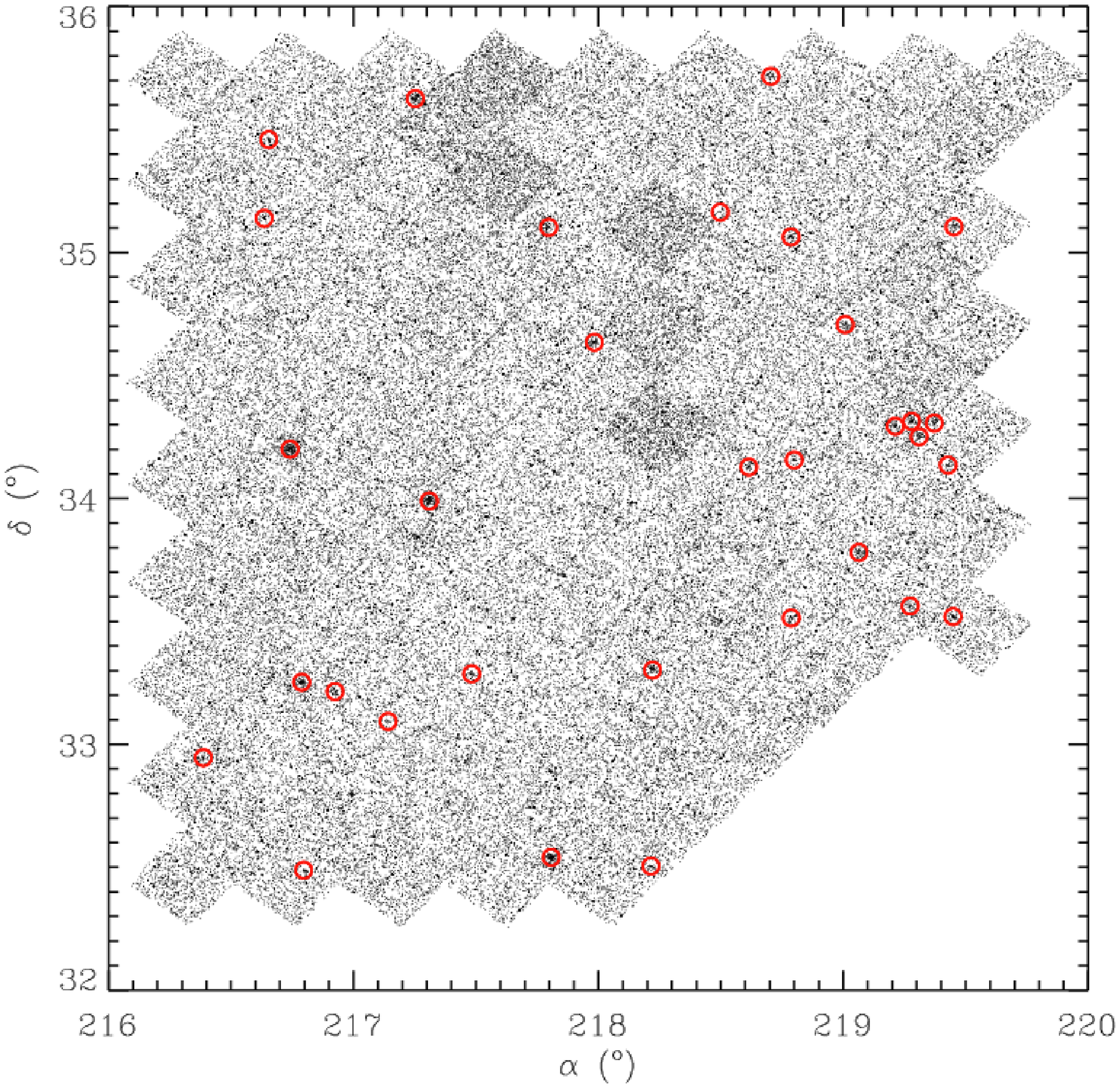}
\end{center}
\caption{Positions of the 32 galaxy groups (red circles) in the Chandra mosaic of Bo\"otes ACIS-I fields}
\label{cand_pos}
\end{figure*}

\par The X-Bo\"otes survey \citep{2005Murray} covers a 9.3 deg$^2$ area of the Bo\"otes constellation (the north field of the NOAO Deep Wide Field Survey), centered on (J2000) R.A. $\approx$ 14:32:00 and Dec $\approx$ +35:06:00. The survey comprises 126 separate contiguous ACIS-I observations each approximately 5 ks in duration. This is the largest mosaic observed by Chandra and allows the study of large-scale structure with arcsecond angular resolution and uniform coverage (e.g. \citet{2011Starikova}).

\par Using a wavelet decomposition, \citet{2005Kenter} detected 3293 point sources in the full 0.5-7 keV band with n $\geq$ 4 X-ray counts.  In addition, they detected 41\footnote{The original extended source list from \citet{2005Kenter} contains 43 objects. But, after determining their redshift, we verified that two sources were identified twice.  They are XBS 01 and XBS 02 and XBS 16 and XBS 17.} extended sources at a existence threshold equivalent to $\approx$ 3$\sigma$, using 0.5-2 keV band images. Due to the volume surveyed, most of these extended sources are expected to be galaxy groups and poor clusters.

\par Additionally we include two extended sources selected by Michael Anderson in 2006 in his SAO REU\footnote{Research Experience for Undergraduate students} summer project on X-ray bright optically normal galaxies (XBONGS) in the Bo\"otes field.  Those sources were selected originally as point sources by \citet{2005Kenter}.  Those two sources together with the 41 detected by \citet{2005Kenter} compose our X-ray sample to generate a catalog of galaxy groups in the Bo\"otes field.

\par In addition to the 5 ks X-ray mosaic, in our analysis we also used all deeper Chandra observations of the Bo\"otes field to confirm that extended sources were not multiple point sources and to better characterize the group properties.  We have 76 ACIS-I observations with 10 to 40 ks of exposure and 14 ACIS-S observations with 10 to 100 ks. The total Chandra exposure time is 1.2 Ms for the Bo\"otes region.

\par To test that each object in the sample is a real galaxy group, we inspected the X-ray, optical and radio images of all 43 candidates and ran the MARX\footnote{http://space.mit.edu/cxc/marx-4.5/index.html} simulations to determine how a point source with the same flux as the real source would appear at the same position in the detector. All simulated sources were compared with the real sources. The profile of the simulated and real sources were fitted to a Gaussian with the width as a free parameter. If the width fitted to the simulated source is consistent within $3\sigma$ or larger than the width fitted to the real one, we confirm it is indeed a point source.  Figure \ref{ps_example} shows three examples of sources originally selected as extended, which we classify as point sources. On the other hand, if the extension of the real source is larger than the simulated source, then it is classed as an extended source.  Figure \ref{es_example} shows an example of an extended source.  After this procedure, we have 32 extended sources that form our galaxy group catalog.

\par Table \ref{tab_sample} provides the general information for the 32 galaxy groups: the group name; coordinates; Chandra ObsIDs; exposure time; final redshift and uncertainties adopted in this work; the number of galaxies used to estimate the redshift and the technique applied  for determining the redshift ($\S$3.1).  Figure \ref{cand_pos} shows the positions of the 32 candidates in the X-Bo\"otes field.

\begin{deluxetable*}{lccccccc}
%\tabletypesize{\tiny}
\tablecaption{Information of position, redshift, and exposure for the 32 galaxy groups selected with the X-Bo\"otes Survey.}
\tablewidth{0pt} 
\tablehead{ 
\colhead{Name} &
\colhead{R.A.} &
\colhead{Dec.} &
\colhead{ObsId} &
\colhead{Exposure} &
\colhead{z} &
\colhead{N$_{spec}^{\rm a}$} &
\colhead{Technique$^{\rm b}$} \\
\colhead{} &
\colhead{(J2000)} &
\colhead{(J2000)} &
\colhead{} &
\colhead{(ks)} &
\colhead{} &
\colhead{} &
\colhead{}
}
\startdata
XBS 02 & 14:25:32.90 & 32:56:44 & 4268     & 4.7  &  0.215 $\pm$ 0.001&  7 & Gap (60")	 \\
       &             &          & 8456     & 5.1  &                   &    &	 \\  
XBS 04 & 14:26:32.51 & 35:08:21 & 3621     & 4.7  &1.75$^{\rm f}$ &  0 & Spectrum \\
       &             &          & 7381     & 4.6	&  		    &    &	 \\  
XBS 05 & 14:26:37.04 & 35:27:34 & 7775$^{\rm c}$ & 14.8 &  0.257            &  1 & BGG$^{\rm d}$ \\
       &             &          & 3605     & 4.7	&  	 	    &    &	 \\  
       &             &          & 7002     & 4.4  &  		    &    &	 \\  
XBS 06 & 14:26:57.90 & 34:12:01 & 10495$^{\rm c}$& 30.0 &  0.129 $\pm$ 0.004& 17 & Gap (60")	 \\
       &             &          & 4224     & 4.7	&  		    &    &	 \\  
       &             &          & 7945     & 40.7	&  		    &    &	 \\  
XBS 07 & 14:27:09.30 & 33:15:10 & 9895$^{\rm c}$ & 30.6 &  0.011            &  1 & BGG$^{\rm d}$ \\
       &             &          & 4255     & 5.0	&  		    &    &	 \\  
XBS 08 & 14:27:13.78 & 32:28:57 & 7948$^{\rm c}$ & 42.2 &  0.132            &  1 & BGG$^{\rm d}$ \\
       &             &          & 4281     & 5.0	&  		    &    &	 \\  
XBS 09 & 14:27:41.89 & 33:12:52 & 4258     & 4.6  &  0.151 $\pm$ 0.000$^{\rm e}$& 13 & Gap (60") \\
XBS 11 & 14:29:00.60 & 35:37:34 & 3602     & 4.5  &  0.234 $\pm$ 0.001&  9 & Gap (60")	 \\
       &             &          & 7000     & 4.4	&  		    &    &	 \\  
       &             &          & 7942     & 38.2	&  		    &    &	 \\  
XBS 13 & 14:29:16.15 & 33:59:29 & 9434$^{\rm c}$ & 24.8 &  0.129 $\pm$ 0.002& 13 & Gap (60")	 \\
       &             &          & 10450    & 22.7	&  		    &    &	 \\  
       &             &          & 4228     & 4.5	&  		    &    &	 \\  
       &             &          & 6983     & 4.7	&  		    &    &	 \\  
       &             &          & 6997     & 9.6  &  		    &    &	 \\                
XBS 14 & 14:29:55.87 & 33:17:11 & 4253     & 4.6  &  0.419 $\pm$ 0.002&  3 & Gap (60")	 \\
XBS 17 & 14:31:09.17 & 35:06:09 & 3624     & 4.6  &  0.194 $\pm$ 0.003& 11 & Gap (60")	 \\
       &             &          & 7378     & 4.4	&  		    &    &	 \\  
XBS 18 & 14:31:13.81 & 32:32:25 & 10496$^{\rm c}$& 28.8 &  0.231 $\pm$ 0.018&  0 & Spectrum\\
       &             &          & 9272     & 5.0	&  		    &    &	 \\  
XBS 20 & 14:31:56.12 & 34:38:06 & 9896$^{\rm c}$ & 50.9 &  0.350 $\pm$ 0.002& 11 & Gap (60")	 \\
       &             &          & 3648     & 4.6	&  		    &    &	 \\  
XBS 21 & 14:32:51.50 & 32:30:18 & 4277     & 4.7  &  ----------       &  0 & ----	 \\
XBS 22 & 14:32:53.14 & 33:18:06 & 4246     & 4.7  &  0.569 $\pm$ 0.000$^{\rm e}$&  5 & Gap (60") \\
       &             &          & 4252     & 4.6	&  		    &    &	 \\  
XBS 25 & 14:34:27.43 & 34:07:46 & 4220     & 4.6  &  0.191 $\pm$ 0.002&  4 & Gap (60")	 \\
       &             &          & 7383     & 4.4	&  		    &    &	 \\  
XBS 26 & 14:34:49.05 & 35:43:01 & 3598     & 4.7  &  0.152 $\pm$ 0.001&  4 & Gap (60")	 \\
XBS 27 & 14:35:08.85 & 35:03:49 & 9435$^{\rm c}$ & 44.6 &  0.730 $\pm$ 0.066&  0 & Spectrum\\
       &             &          &  3626    &  4.7	&  		    &    &	 \\  
       &             &          & 7376     & 4.4	&  		    &    &	 \\  
XBS 28 & 14:35:09.03 & 33:30:50 & 4247     & 4.7  &  0.422 $\pm$ 0.138&  1 & BGG$^{\rm d}$ \\
       &             &          & 7011     & 4.7	&  		    &    &	 \\  
XBS 29 & 14:35:11.94 & 34:09:22 & 13132$^{\rm c}$& 27.7 &  0.404            &  1 & BGG$^{\rm d}$ \\
       &             &          & 4219     & 4.7	&  	            &    &	 \\  
       &             &          & 7383     & 4.4	&  		    &    &	 \\  
XBS 32 & 14:36:01.94 & 34:42:26 & 3643     & 4.7  &  0.534 $\pm$ 0.001&  4 & Gap (60")	 \\
       &             &          & 7003     & 4.4	&  		    &    &	 \\  
XBS 33 & 14:36:15.44 & 33:46:50 & 4232     & 4.7  &  0.343 $\pm$ 0.001&  6 & Gap(60'')	 \\
       &             &          & 6979     & 5.1	&  		    &    &	 \\  
XBS 35 & 14:36:51.06 & 34:17:37 & 3659     & 4.7  &  0.045            &  1 & BGG$^{\rm d}$ \\
       &             &          & 7382     & 4.8	&  	            &    &	 \\  
XBS 36 & 14:37:05.56 & 33:33:44 & 4234     & 4.7  &  0.243 $\pm$ 0.000$^{\rm e}$&  4 & Gap (60") \\
XBS 37 & 14:37:07.06 & 34:18:48 & 3659     & 4.7  &  0.122 $\pm$ 0.001&  5 & Gap (60")	 \\
XBS 38 & 14:37:14.35 & 34:15:03 & 4218     & 5.0  &  0.541 $\pm$ 0.002&  8 & Gap(60'')	 \\
XBS 39 & 14:37:29.18 & 34:18:22 & 3660     & 4.6  &  0.396            &  1 & BGG$^{\rm d}$ \\
XBS 41 & 14:37:42.77 & 34:08:07 & 10461$^{\rm c}$& 100.0&  0.543 $\pm$ 0.186&  1 & BGG$^{\rm d}$ \\
       &             &          & 4218     & 5.0	&  		    &    &	 \\  
XBS 42 & 14:37:47.63 & 33:31:10 & 4249     & 4.6  &  0.218 $\pm$ 0.021&  1 & BGG$^{\rm d}$ \\
XBS 43 & 14:37:48.49 & 35:06:17 & 3628     & 4.6  &  0.574 $\pm$ 0.002&  3 & Gap (0.5 Mpc)	 \\
XBS 46 & 14:28:33.80 & 33:05:35 & 4258     & 4.6  &  0.196 $\pm$ 0.000$^{\rm e}$&  3 & Gap (60") \\
XBS 52 & 14:43:35.94 & 35:09:51 & 3615     & 4.6  &  0.599 $\pm$ 0.015&  1 & BGG$^{\rm d}$	 \\
       &             &          & 7376     & 4.4	&                   &    &    	 \\
 \enddata
\tablenotetext{a}{Number of galaxy redshifts used to estimate the group redshift}
\tablenotetext{b}{Technique applied to estimate the group redshift}
\tablenotetext{c}{ACIS-S observations}
\tablenotetext{d}{Brightest Galaxy of the Group}
\tablenotetext{e}{Redshift uncertainties $ < 0.001$}
\tablenotetext{f}{This redshift was estimated by \citet{2012Stanford}}
\label{tab_sample}
\end{deluxetable*}

\subsection{AGN and Galaxy Evolution Survey (AGES)}
\label{ages}

\par The AGN and Galaxy Evolution Survey (AGES, \citet{2012Kochanek}) is a redshift survey, covering 7.88 deg$^2$ of the Bo\"otes field. The observations were made with the Hectospec instrument, a robotic spectrograph with 300 fibers in a one degree field of view on the 6.5 m MMT telescope.  Each fiber has a diameter of 1.5 arcseconds. The spectrograph is positioned in 15 different locations on the sky as shown in Figure 2 from \citet{2012Kochanek}.  The wavelength range is 3700 \AA~ to 9200 \AA, with a pixel scale of 1.2 \AA ~ and a spectral resolution of 6 \AA.  The AGES survey was designed to investigate the AGN activity and properties of galaxies from the local to the distant Universe.

\par The final sample comprises 21,805 redshifts for galaxies and AGNs to a limiting magnitude of I $<$ 20 mag, having 4764 of 21,805 AGN spectra.  The survey is sensitive to L* galaxies to $z = 0.5$.  The median galaxy redshift is 0.31 and 90\% of the redshifts are in the range 0.05 $< z <$ 0.66.

%%%%%%%%%%%%%%%%%%%%%%%%%%%%%%%%%%%%%%%%%%%%%%%%%%%%%%%%%%%%%%%%%%%%%%%%
%%%%%%%%%%%%%%%%%%%%%%%%%%%%%%%%%%%%%%%%%%%%%%%%%%%%%%%%%%%%%%%%%%%%%%%%
%%%%%%%%%%%%%%%%%%%%%%%%%%%%%%%%%%%%%%%%%%%%%%%%%%%%%%%%%%%%%%%%%%%%%%%%
%%%                                                                  %%%
%%%             Dynamical Analysis of Galaxy Groups                  %%%
%%%                                                                  %%%
%%%%%%%%%%%%%%%%%%%%%%%%%%%%%%%%%%%%%%%%%%%%%%%%%%%%%%%%%%%%%%%%%%%%%%%%
%%%%%%%%%%%%%%%%%%%%%%%%%%%%%%%%%%%%%%%%%%%%%%%%%%%%%%%%%%%%%%%%%%%%%%%%
%%%%%%%%%%%%%%%%%%%%%%%%%%%%%%%%%%%%%%%%%%%%%%%%%%%%%%%%%%%%%%%%%%%%%%%%

\section{Dynamical Analysis of Galaxy Groups}
\subsection{Redshift Determinations for Each Group}
\label{redshift}

\begin{figure}[!htb]
\begin{center}
\includegraphics[width=80mm]{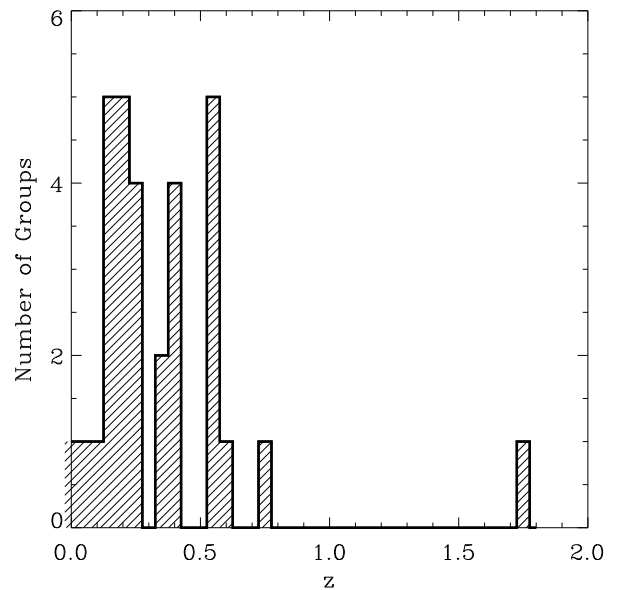}
\end{center}
\caption{The redshift distribution of the 30 X-ray selected groups in the Bo\"otes field}
\label{z_dist}
\end{figure}

\par We use the AGES data to determine the redshifts of the galaxy groups.  Each system's redshift was first estimated inside a 60 arcsecond radius aperture centered on the X-ray position. We use two different approaches to estimate the galaxy group's redshift.  In the first, we try to identify each group in redshift space by applying a "gap-technique" \citep{1996Katgert}.  If we identify the group in redshift space, then we measure its redshift applying a biweight estimate, using only those galaxies selected by the gap-technique.  If we do not identify a group in redshift space, we apply a second approach which consists of determining the redshift of the group applying the biweight estimate to all galaxies within 60 arcseconds of the X-ray position, no matter their redshift differences.

\par The gap-technique \citep{1996Katgert} identifies each group in redshift space.  Two galaxies, which are adjacent in redshift, are determined to belong to the same group, if their velocity difference does not exceed a pre-determined value, called the velocity gap.  We adopt a variable velocity gap referred to as a 'density gap' \citep{1998Adami, 2007Lopes, 2009aLopes}.  The density gap width is given by the expression

\begin{equation}
\Delta z = 500\{1 + {\rm exp}[-(N-6)/33]\}/c.
\end{equation}

\noindent where \textit{N} is the number of galaxies inside the 60 arcsecond radius aperture and \textit{c} is the speed of light in km s$^{-1}$.  The gap-technique considers the system to be a galaxy group, if it has at least three galaxies identified.  After a galaxy group is identified through the gap-technique, we determine its redshift through the biweight estimate \citep{1990Beers}, using only the galaxies selected with the gap-technique. If more than one group of galaxies is identified, we choose the group with the smallest offset from the X-ray centroid. After the redshift determination, we diagnosed two groups that were identified twice in X-rays by \citet{2005Kenter}. They are groups XBS 01 and XBS 02, and XBS 16 and XBS 17.  For our analysis, we keep the groups XBS 02 and XBS 17, because they have more galaxies identified as members compared to their duplicate identifications.

\par With the gap-technique, we are able to identify 17 of 32 extended sources in redshift space.  For the other 15 systems, the biweight estimate is performed, using all galaxies within 60 arcseconds of the X-ray center.  If there is only one galaxy with a measured redshift, its redshift is taken as the redshift of the system.  Using this second approach, we are able to estimate the redshifts for 10 of the 15 groups.  Since the gap-technique combined with the biweight estimate provides more reliable results than just the biweight estimate, we attempt to refine the redshifts determined from the biweight analysis.  To do this, we compute the 0.5 Mpc radius using the redshift estimated from the biweight analysis and we apply the gap-technique inside this physical aperture.  If we identify a group in redshift space within 0.5 Mpc, we perform a biweight estimate using the galaxies now identified to be associated with the group.  With this new approach, we are able to determine redshifts for 5 of the 10 groups with redshifts derived only through the biweight estimate.  Finally there are five groups which do not have AGES data (XBS 04, 08, 18, 21 and 27).  For XBS 04, we adopted the redshift determined from optical and IR spectra by \citet{2012Stanford}. This cluster was originally identified using the SDWFS (Spitzer Deep Wide Field Survey,\citet{2009Ashby}) data and matched with the NDWFS data. There are seven members spectroscopically identified, one of them is a QSO observed in AGES.  For XBS 08, we adopted the redshift of the BCG from the Sloan data as the group redshift. For XBS 18 and 27, we determined the redshift from the X-ray spectrum.  XBS 21 has no redshifts. The list of the groups with their respective redshifts and the technique applied is given in Table 1.  

\par To assure the reliability of the group redshifts (especially the groups XBS 05, 28, 29, 39, 41 and 52 for which the redshift was estimated based on only one or two galaxies), we identify the brightest galaxy of the group (BGG) with $M_R$ $\leq$ $M^*_R - 1$, inside an aperture of 60 arcseconds centered on the X-ray emission and compare its redshift with the redshift of the group.  When the redshift difference $|cz_{group}-cz_{BGG}|$ $\geq$ 300 km s$^{-1}$, we adopt the redshift of the BGG as the redshift of the system.  We could identify the BGG in 28 of the 32 groups. For these 28 groups, 16 have a group redshift consistent with the redshift of the brightest galaxy of the group, 6 only have the BGG redshift and the remaining 6 groups do not have a redshift consistent with the BGG redshift. For these 6 groups we adopt the redshift of the BGG as the redshift of the group. Table \ref{tab_bcg} shows the information for the BGGs.  Thus, the final group sample contains 32 X-ray selected groups and we are able to determine the redshifts for 31 of these.

\begin{deluxetable}{lccccc}
\tablecaption{Information of the Brightest Galaxy of the Group}
\tablewidth{0pt} 
\tablehead{ 
\colhead{Name} &
\colhead{R.A.$_{BGG}$} &
\colhead{Dec$_{BGG}$} &
\colhead{z$_{BGG}$} &
\colhead{$M_R$} &
\colhead{Offset$^{\rm a}$} \\
\colhead{} &
\colhead{(deg)} &
\colhead{(deg)} &
\colhead{} &
\colhead{} &
\colhead{(arcmin)}
}
\startdata
XBS 02&      216.387&      32.9440&     0.214&       -22.62&     0.11\\ %&      23\\
XBS 05&      216.655&      35.4588&     0.257&       -21.53&     0.06\\ %&      15\\
XBS 06&      216.749&      34.1999&     0.128&       -23.90&     0.36\\ %&      49\\
XBS 07$^{\rm b}$&      216.786&      33.2525&     0.011&       -20.51&     0.13\\ %&      2\\
XBS 08$^{\rm c}$&      216.797&      32.4880&     0.132&       -23.19&     0.68\\ %&      96\\
XBS 09&      216.931&      33.2042&     0.151&       -21.83&     0.71\\ %&      112\\
XBS 11&      217.251&      35.6226&     0.234&       -23.15&     0.22\\ %&      49\\
XBS 13&      217.310&      33.9896&     0.129&       -23.85&     0.39\\ %&      55\\
XBS 14&      217.482&      33.2864&     0.419&       -23.78&     0.04\\ %&      12\\
XBS 17&      217.798&      35.1026&     0.191&       -22.71&     0.46\\ %&      87\\
XBS 20&      217.980&      34.6349&     0.349&       -23.36&     0.24\\ %&      71\\
XBS 22&      218.220&      33.3062&     0.569&       -24.50&     0.26\\ %&      103\\
XBS 25&      218.615&      34.1279&     0.189&       -22.93&     0.11\\ %&      20\\
XBS 26&      218.705&      35.7131&     0.152&       -20.88&     0.01\\ %&      21\\
XBS 28&      218.788&      33.5182&     0.422&       -23.49&     0.25\\ %&      83\\
XBS 29&      218.810&      34.1496&     0.404&       -20.28&     0.54\\ %&      174\\
XBS 32&      219.019&      34.6994&     0.534&       -24.23&     0.72\\ %&      271\\
XBS 33&      219.057&      33.7881&     0.337&       -23.49&     0.51\\ %&      146\\
XBS 35&      219.211&      34.2927&     0.045&       -21.47&     0.09\\ %&      5\\
XBS 36&      219.271&      33.5610&     0.243&       -23.66&     0.11\\ %&      25\\
XBS 37&      219.279&      34.3140&     0.122&       -23.59&     0.03\\ %&      4\\
XBS 38&      219.319&      34.2516&     0.547&       -24.14&     0.33\\ %&      125\\
XBS 39$^{\rm c}$&      219.380&      34.3097&     0.396&       -19.75&     0.45\\ %&      144\\
XBS 41&      219.429&      34.1363&     0.543&       -24.28&     0.07\\ %&      24\\
XBS 42&      219.448&      33.5205&     0.218&         ----&     0.06\\ %&      13\\
XBS 43&      219.442&      35.1138&     0.576&       -23.37&     0.65\\ %&      255\\
XBS 46&      217.141&      33.0890&     0.196&       -22.66&     0.03\\ %&      8\\
XBS 52&      218.498&      35.1645&     0.599&       -22.63&     0.02\\ %&      9\\
 \enddata
\tablenotetext{a}{Offset between the X-ray position and the BGG}
\tablenotetext{b}{Redshift from NASA/IPAC Extragalactic Database (NED)}
\tablenotetext{c}{Redshift from SDSS DR8}
\label{tab_bcg}
\end{deluxetable}

\begin{figure*}[!htb]
\begin{center}
\includegraphics[width=54mm]{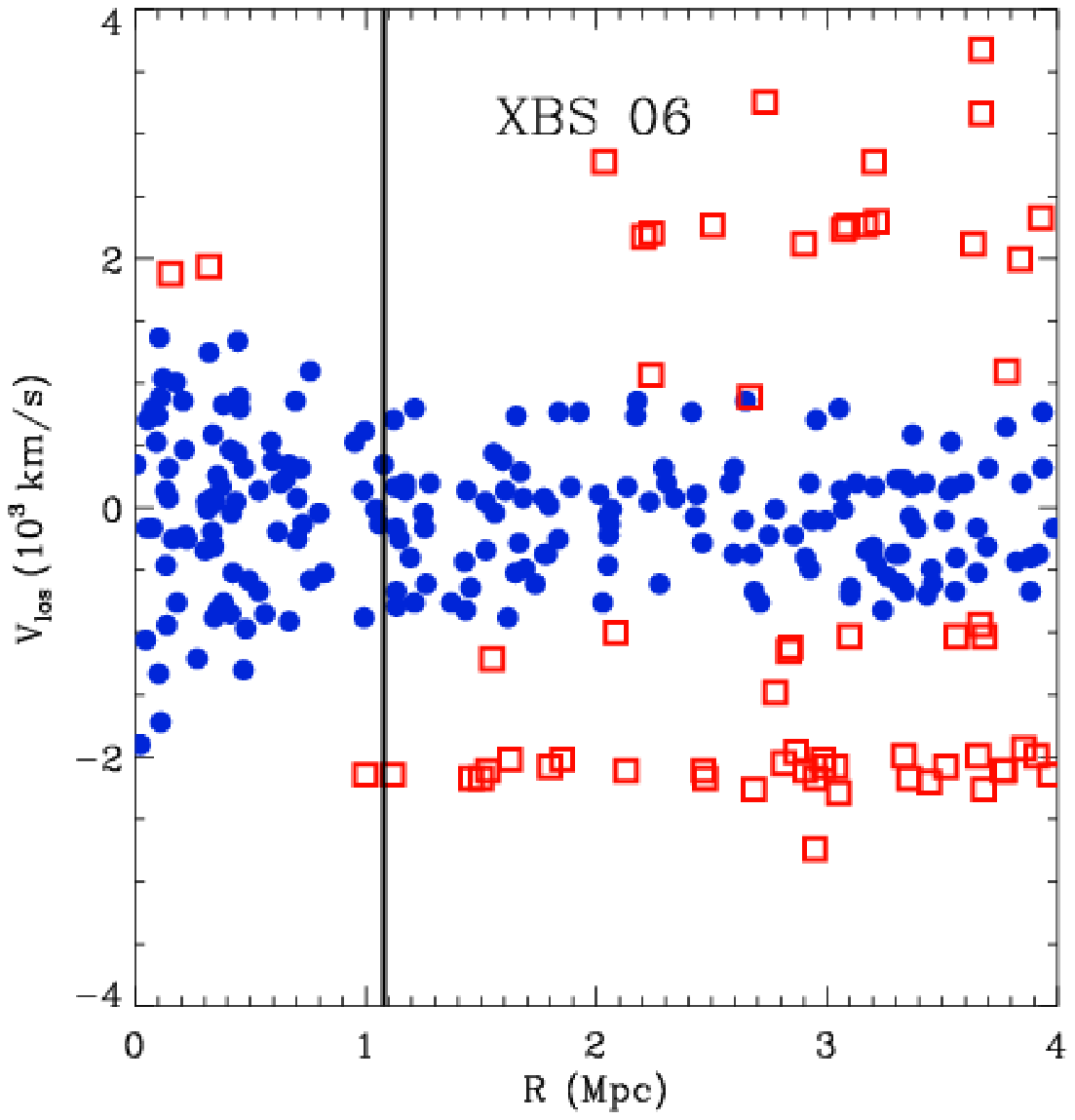}
\includegraphics[width=54mm]{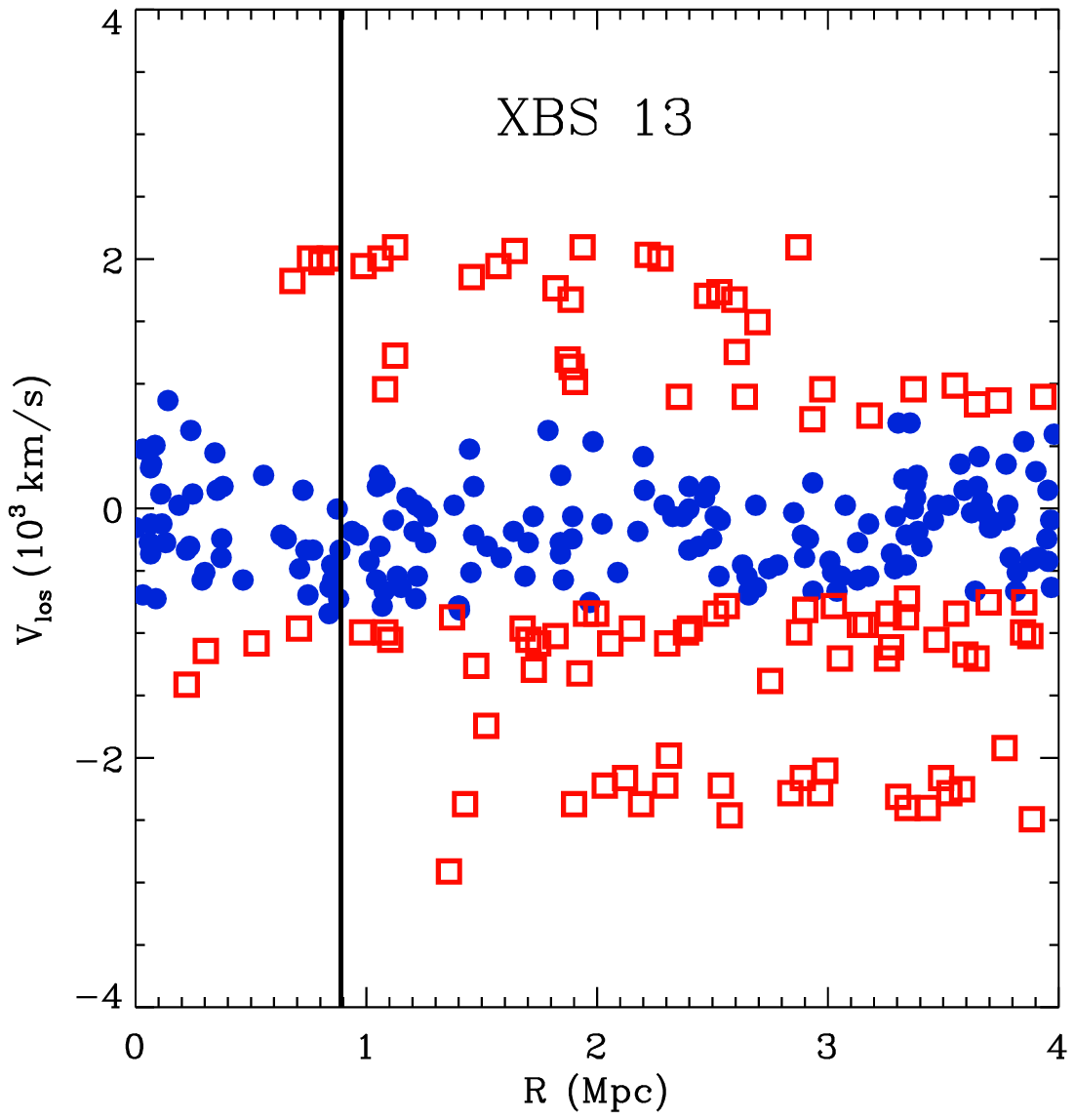}
\includegraphics[width=54mm]{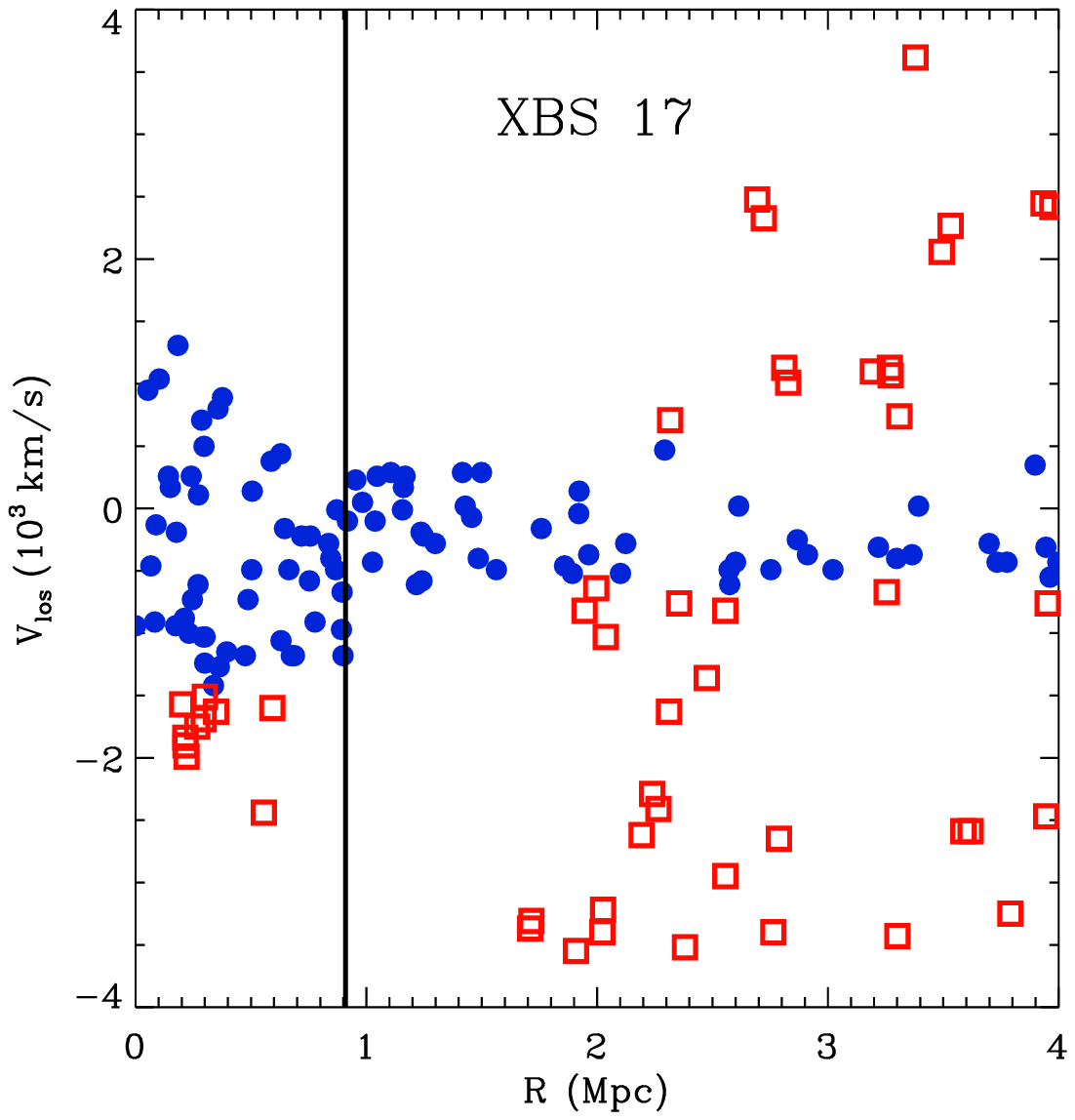}
\end{center}
\caption{Phase-space diagrams of 3 galaxy groups shown as examples. The velocity and radial offsets are with respect to the group center. We apply a shifting gapper procedure for the selection of group members (filled blue circles) and exclusion of interlopers (open red squares).  The vertical line is the R$_{500}$ of the group.  The large number of members at large radii sensitive to the group velocity illustrate the large structure in which these groups reside}
\label{velrad}
\end{figure*}

\subsection{Virial Analysis}
\par With the redshifts estimated, we calculate the dynamical properties of the groups, including velocity dispersion, physical radius and total mass. To do this, we perform a virial analysis. Lopes et al. (2009a) showed that if a survey is complete to at least M$^*$+1, the velocity dispersion ($\sigma_{gr}$) and the corresponding mass estimates for galaxy groups and clusters are reliable, while if the survey is shallower, those masses may be biased. Therefore, we apply the virial analysis only for the groups with $z \leq$ 0.35, where AGES is complete to M$^*$+1.  We also require that the galaxy group must have at least five galaxy members.

\par The first step before starting the virial analysis is to eliminate the interlopers.  To do so, we use a technique called ``shifting-gapper'' \citep{2009aLopes}, applied to all galaxies with spectra available within $R_{500}$ centered on the X-ray coordinates. The $R_{500}$ of each group is determined from the $L_X-M_{500}$ relation for groups from \citet{2011Eckmiller}, using the X-ray luminosities estimated in Section $\S$\ref{xray_prop}. We convert $M_{500}$ to $R_{500}$ using the equation:

\begin{equation}
M_{500} = 500\times \frac{4\pi}{3}\rho_c R_{500}^3,
\end{equation}

\noindent where $\rho_c$ is the critical density of the Universe at the redshift of the group.

\par The principle of the ``shifting-gapper'' technique is the same as for the gap technique used to estimate the redshifts of the groups, except for two differences: (i) we apply the procedure in radial bins from the center of the group; (ii) the size of the velocity gap between two galaxies adjacent in velocity depends on the velocity limits of each group (as shown in Equation 3).  This method makes no assumption about the dynamical state of the group.  The procedure applied is very similar to the one used in \citet{2009aLopes}.  The most important difference is that we do not run the interloper rejection to a radius of  4 Mpc, but instead we stop at $R_{500}$.  The reason why we perform the analysis only inside $R_{500}$ is because for poor groups, 4 Mpc is several times larger than the virial radius.

\par We start the ``shifting-gapper''technique by selecting all galaxies within $R_{500}$ of the group center and with $|cz - cz_{group}|$ $\leq$ 4000 km s$^{-1}$, where $z_{group}$ is the redshift of the group estimated in $\S$3.1.  We run the gap technique in radial bins of 0.35 Mpc width or larger (if it does not exceed $R_{500}$) to guarantee the selection of at least 15 galaxies  in each annulus.  In the first radial bin, we order the galaxies by velocity and run the gap technique.  A galaxy is excluded as an outlier, if the velocity difference between it and the adjacent galaxy exceeds the velocity gap given by:

\begin{equation}
\Delta z = \frac{\lvert v_{lo} - v_{hi}\rvert}{10} \times \{1 + {\rm exp}[-(N - 6)/3]\}/c.
\label{shifting-gap_eq}
\end{equation}

\noindent where $v_{lo}$ and $v_{hi}$ are, respectively, the lower and higher velocity limits of the galaxies in the group measured inside 0.5 Mpc. These velocity limits are symmetric.

\par If no group is found in the first radial bin, we keep all the galaxies, $v_{lo}$ and $v_{hi}$ and we reanalyze them with the galaxies of the next annulus.  If a group is identified according to Eq. \ref{shifting-gap_eq}, we keep all the galaxies selected as members within $v_{lo}$ and $v_{hi}$.  Then, we update the values of $v_{lo}$ and $v_{hi}$, using only the member galaxies.  For the next annulus, we use these new values for $v_{lo}$ and $v_{hi}$.  We repeat this routine until we reach R$_{500}$.

\par After excluding the interlopers, we have 14 groups within $z \leq 0.35$ with at least 5 galaxy members.  Figure \ref{velrad} illustrates the results of the outliers removal for the three most massive groups in the sample.  In each panel, the filled circles to the left of the solid line represent the group members and the open squares represent the rejected interlopers.  The vertical line marks $R_{500}$ for each group from the $L_X$-$M_{500}$ relation \citep{2011Eckmiller}.  In Figure \ref{velrad}, we extended the interloper rejection analysis to larger radii to show the large scale structure in which these groups reside.  The result of the outlier exclusion inside $R_{500}$ is not altered, if we extend the analysis to large radii.

\par In Appendix A, we present the galaxy group catalog to $z \leq 0.35$.  We show the galaxy number density profile and the spatial distribution of galaxy members and interlopers for each of these 14 groups.  In the numeric density profiles (left panel of Figure \ref{finding_charts}), we can see that even the least massive groups present a peak of density over the galaxy background distribution, reinforcing that these systems are real galaxy groups.  In the spatial distribution of galaxy members and interlopers(middle and right panels of Figure \ref{finding_charts}), we also see a clump of galaxy members inside $R_{500}$ for all 14 groups.  The colors of the galaxy members and interlopers represent their velocities.  Through the colors of the symbols, it is possible to verify that these central galaxy clumps are not a projection effect, but a real agglomeration of galaxies at the same distance.  

\par These 14 groups were subjected to the virial analysis adopted by \citet{1998Girardi}, \citet{2005Popesso}, \citet{2007Popesso}, \citet{2006Biviano} and \citet{2009aLopes}.  For more details about the rejection of interlopers and the virial analysis procedures see \citet{2009aLopes}.  As a result of this procedure, we obtain $\sigma_{gr}$, $R_{500}$, $R_{200}$, $M_{500}$ and $M_{200}$. In our further analysis, we adopt the $R_{500}$ estimated through the virial analysis. The results of the virial analysis are given in Table \ref{tab_opt}: the group name; number of galaxies used to compute the velocity dispersion ($N_{\sigma}$); velocity dispersion ($\sigma_{gr}$); physical radius ($R_{500}$); total mass ($M_{500}$); optical luminosity ($L_{opt}$) and richness ($N_{gals}$) - those final two quantities will be derived in Sec. $\S$\ref{opt_prop}.  The X-Bo\"otes groups have velocity dispersion estimates of 110 $<$ $\sigma_{gr}$ $<$ 650 km s$^{-1}$.  We find that these 14 groups have total masses which range over two magnitudes from 3$\times$10$^{12}$ M$_{\odot}$ to 2$\times$10$^{14}$ M$_{\odot}$.

%%%%%%%%%%%%%%%%%%%%%%%%%%%%%%%%%%%%%%%%%%%%%%%%%%%%%%%%%%%%%%%%%%%%%%%%
%%%%%%%%%%%%%%%%%%%%%%%%%%%%%%%%%%%%%%%%%%%%%%%%%%%%%%%%%%%%%%%%%%%%%%%%
%%%%%%%%%%%%%%%%%%%%%%%%%%%%%%%%%%%%%%%%%%%%%%%%%%%%%%%%%%%%%%%%%%%%%%%%
%%%                                                                  %%%
%%%                      Optical Properties                          %%%
%%%                                                                  %%%
%%%%%%%%%%%%%%%%%%%%%%%%%%%%%%%%%%%%%%%%%%%%%%%%%%%%%%%%%%%%%%%%%%%%%%%%
%%%%%%%%%%%%%%%%%%%%%%%%%%%%%%%%%%%%%%%%%%%%%%%%%%%%%%%%%%%%%%%%%%%%%%%%
%%%%%%%%%%%%%%%%%%%%%%%%%%%%%%%%%%%%%%%%%%%%%%%%%%%%%%%%%%%%%%%%%%%%%%%% 

\section{Optical Properties}
\label{opt_prop}

\par For each of the 14 groups with mass estimates, we derive the richness and optical luminosity. Using the photometric data from the NDWFS,  we apply the same procedure used in \citet{2006Lopes,2009aLopes}.  Our richness (N$_{gals}$) is defined as the number of galaxies with $m^{*}_{R}-1 \leq m_{R} \leq m^{*}_{R}+2$ inside R$_{500}$, where $m^{*}_{R}$ is the characteristic apparent magnitude in R band of the cluster luminosity function.  Since the NDWFS is complete to $M^*+2$ to $z=1.37$, we guarantee that no galaxy is lost between the magnitude range considered to estimate the optical richness.  The optical luminosity was also estimated inside $R_{500}$.  We adopt the values for the bright end of the Schechter luminosity function, obtained by \citet{2006Popesso}.  They found the slope of the bright end $\alpha = -1.09$ and $M^* = -20.94$ inside $R_{200}$.  We convert the values of $M^*$ for $z = 0$ using the cosmology adopted in this work.  For different redshifts we apply an evolutionary correction for $M^*$ from \citet{1999Yee}:

\begin{equation}
M^*(z) = M^*(0) -Qz,
\end{equation}

\noindent where $Q = -1.4$.

\par The first step in estimating the group richness is to convert $M^*$ into $m^*$ and calculate the apparent radius in arcseconds for $R_{500}$.  As we are only interested in galaxies with magnitudes between $m^{*}_{R}-1 \leq m_R \leq m^{*}_{R}+2$, we selected all objects inside $R_{500}$ with magnitudes between $m^{*}_{R} - 1 + k_s \leq m_R \leq m^{*}_{R} + 2 + k_e$, where $k_e$ and $k_s$ are, respectively, the $k$ correction for elliptical and spiral (Sbc) galaxies.  Considering the $k$ correction in the magnitude limits, we can guarantee that all galaxies inside $m^{*}_{R}-1 \leq m_R \leq m^{*}_{R}+2$ are selected.  We define $N_{grp}$ as the number of galaxies inside $R_{500}$ and within the magnitude range defined previously.

\par Next we estimate the galaxy background contribution and subtract it from $N_{grp}$.  We used 20 fields each $0.5^\circ$ in aperture spread randomly inside a $1^\circ$ annulus, 3 Mpc distant from the center of each galaxy group.  In each field, we counted the number of galaxies within the same magnitude range used to extract $N_{grp}$.  The background contribution of galaxies ($N_{bkg}$) is given by the median number in the 20 fields.  To avoid the border effect and contamination by other systems, fields with galaxy counts higher than $3\sigma$ or lower than $2\sigma$, were excluded from the median.  Finally,  the corrected galaxy counts in the group, $N_{cor}$, is equal to $N_{grp}-N_{bkg}$, where $N_{bkgd}$ is normalized by the source area.

\par Next, we apply the $k$ correction to the galaxies inside $R_{500}$, following a bootstrap procedure.  This method consists of randomly selecting $N_{cor}$ galaxies from $N_{grp}$ galaxies and applying the $k$ correction to each one.  An elliptical $k$ correction is applied to $X$ percent of the $N_{cor}$ galaxies, while a Sbc $k$ correction is applied to $100-X$ percent.  This percentage $X$ of $k_e$ depends on the redshift.  We assume that at $z \leq 0.15$ galaxy clusters and groups are composed of 80\% early-type galaxies (E and S0), while at $0.15 < z \leq 0.30$ the percentage drops to 50\% and for $z > 0.30$, it is equal to 30\% \citep{1997Dressler,2005Smith,2009aLopes}.  Then, with the corrected magnitudes, we can count the number of galaxies in the range $m^{*}_{R}-1 \leq m_R \leq m^{*}_{R}+2$.  We repeat this procedure 100 times.  The final value of $N_{gals}$ is provided by the median of the 100 procedures.  The richness uncertainty ($\sigma_{N_{gals}}$) is a combination of the uncertainties of the background count ($\sigma_{bkg}$) and the bootstrap procedure ($\sigma_{boot}$). Thus, $\sigma_{N_{gals}}$ is equal to $\sqrt{\sigma_{bkg}^{2} + \sigma_{boot}^{2}}$.

\par We applied the same bootstrap procedure to estimate the optical luminosity for each of the 14 groups.  Using bins of $\Delta$mag $= 0.02$, we generated a magnitude distribution for the $N_{cor}$ galaxies inside $R_{500}$ with magnitudes within $m^{*}_{R}-1 \leq m_R \leq m^{*}_{R}+2$.  The total optical luminosity ($L_{opt}$) of each system is given by:

\begin{equation}
L_{opt} = \sum^{n}_{i=1} N_{i} 10^{-0.4R_{i}},
\label{lopt_eq}
\end{equation}

\noindent where $N_i$ is the corrected counts for each magnitude bin $R_i$.  Then, we transformed the optical luminosity to absolute magnitude ($M^{R}_{grp}$) and applied the $k$ correction. To obtain the optical luminosity in solar units, we used the following equation:

\begin{equation}
L_{opt} = 10^{-0.4(M^{R}_{grp} - M^{R}_{\odot})}
\end{equation}
\noindent where $M_{\odot}^{R} = 4.42$ \citep{2007Blanton}.  The results for $L_{opt}$ and $N_{gals}$ inside $R_{500}$ for each group can be found in Table \ref{tab_opt}.  The X-Bo\"otes groups have optical luminosities 10$^{11}$ L$_{\odot}$ $<$ $L_{opt}$ $<$ 10$^{12}$ L$_{\odot}$, with 9 groups ($\sim$ 65\%) having $L_{opt}$ $\leq$ 0.5$\times$10$^{12}$ L$_{opt}$. Richnesses are in the range 5 $<$ $N_{gals}$ $<$ 60. Hence, as expected, the $L_{opt}$ and $N_{gals}$ have values typical of galaxy groups.

\begin{deluxetable}{lcccccc}
\tablecaption{Dynamical and Optical Properties For 14 Groups with $z\leq0.35$ and at least 5 Galaxy Members}
\tablewidth{0pt} 
\tablehead{ 
\colhead{Name} &
\colhead{$N_{\sigma}$} &
\colhead{$\sigma_{gr}$} &
\colhead{$R_{500}$} &
\colhead{$M_{500}$} &
\colhead{$L_{opt}$} &
\colhead{$N_{gals}$} \\
\colhead{} &
\colhead{} &
\colhead{(km s$^{-1}$)} &
\colhead{(Mpc)} &
\colhead{(10$^{14}$ M$_{\odot}$)} &
\colhead{(10$^{12}$ L$_{\odot}$)} &
\colhead{}
}
\startdata
XBS 02&      17& 184$^{+60}_{-25}$&     0.32$^{+0.07}_{-0.03}$&     0.12$^{+0.08}_{-0.03}$&     0.28$\pm$0.11&      15$\pm$0.8\\
       &        &                           &                           &                           &                  &                \\
XBS 06&      81& 641$^{+58}_{-41}$&     0.88$^{+0.05}_{-0.04}$&     2.19$^{+0.39}_{-0.28}$&     1.19$\pm$0.24&      50$\pm$3.3\\
       &        &                           &                           &                           &                  &                \\
XBS 07&       9& 160$^{+52}_{-21}$&     0.33$^{+0.07}_{-0.03}$&     0.10$^{+0.07}_{-0.03}$&     0.11$\pm$0.05&       9$\pm$0.4\\
       &        &                           &                           &                           &                  &                \\
XBS 09&       7& 242$^{+99}_{-18}$&     0.34$^{+0.10}_{-0.03}$&     0.13$^{+0.11}_{-0.03}$&     0.29$\pm$0.12&      14$\pm$0.3\\
       &        &                           &                           &                           &                  &                \\
XBS 11&      22& 417$^{+69}_{-37}$&     0.58$^{+0.06}_{-0.04}$&     0.70$^{+0.23}_{-0.13}$&     0.74$\pm$0.22&      23$\pm$2.1\\
       &        &                           &                           &                           &                  &                \\
XBS 13&      42& 375$^{+48}_{-31}$&     0.59$^{+0.05}_{-0.03}$&     0.65$^{+0.17}_{-0.11}$&     0.32$\pm$0.13&      17$\pm$1.7\\
       &        &                           &                           &                           &                  &                \\
XBS 17&      49& 615$^{+66}_{-49}$&     0.85$^{+0.06}_{-0.05}$&     2.14$^{+0.46}_{-0.35}$&     1.40$\pm$0.27&      59$\pm$2.2\\
       &        &                           &                           &                           &                  &                \\
XBS 25&      11& 377$^{+89}_{-30}$&     0.55$^{+0.09}_{-0.03}$&     0.58$^{+0.28}_{-0.10}$&     0.35$\pm$0.14&      13$\pm$1.2\\
       &        &                           &                           &                           &                  &                \\
XBS 26&      13& 163$^{+43}_{-22}$&     0.39$^{+0.07}_{-0.04}$&     0.20$^{+0.11}_{-0.06}$&     0.17$\pm$0.08&      11$\pm$0.6\\
       &        &                           &                           &                           &                  &                \\
XBS 33&      14& 144$^{+41}_{-18}$&     0.38$^{+0.07}_{-0.03}$&     0.22$^{+0.13}_{-0.06}$&     0.94$\pm$0.22&      34$\pm$0.8\\
       &        &                           &                           &                           &                  &                \\
XBS 35&       7& 123$^{+44}_{-15}$&     0.21$^{+0.05}_{-0.02}$&     0.03$^{+0.02}_{-0.01}$&     0.10$\pm$0.04&       9$\pm$0.2\\
       &        &                           &                           &                           &                  &                \\
XBS 36&       6& 111$^{+74}_{-39}$&     0.23$^{+0.10}_{-0.05}$&     0.04$^{+0.06}_{-0.03}$&     0.23$\pm$0.08&      14$\pm$0.1\\
       &        &                           &                           &                           &                  &                \\
XBS 37&      17& 335$^{+51}_{-27}$&     0.52$^{+0.05}_{-0.03}$&     0.44$^{+0.14}_{-0.08}$&     0.51$\pm$0.17&      21$\pm$0.9\\
       &        &                           &                           &                           &                  &                \\
XBS 46&       7& 129$^{+52}_{-24}$&     0.28$^{+0.08}_{-0.04}$&     0.08$^{+0.06}_{-0.03}$&     0.17$\pm$0.10&       7$\pm$0.2\\
 \enddata
\label{tab_opt}
\end{deluxetable}

%%%%%%%%%%%%%%%%%%%%%%%%%%%%%%%%%%%%%%%%%%%%%%%%%%%%%%%%%%%%%%%%%%%%%%%%
%%%%%%%%%%%%%%%%%%%%%%%%%%%%%%%%%%%%%%%%%%%%%%%%%%%%%%%%%%%%%%%%%%%%%%%%
%%%%%%%%%%%%%%%%%%%%%%%%%%%%%%%%%%%%%%%%%%%%%%%%%%%%%%%%%%%%%%%%%%%%%%%%
%%%                                                                  %%%
%%%                        X-ray Properties                          %%%
%%%                                                                  %%%
%%%%%%%%%%%%%%%%%%%%%%%%%%%%%%%%%%%%%%%%%%%%%%%%%%%%%%%%%%%%%%%%%%%%%%%%
%%%%%%%%%%%%%%%%%%%%%%%%%%%%%%%%%%%%%%%%%%%%%%%%%%%%%%%%%%%%%%%%%%%%%%%%
%%%%%%%%%%%%%%%%%%%%%%%%%%%%%%%%%%%%%%%%%%%%%%%%%%%%%%%%%%%%%%%%%%%%%%%% 

\section{X-ray Properties}
\label{xray_prop}

\begin{deluxetable}{lccc}
\tablecaption{X-ray Surface Brightness Profile Fit Parameters For 11 Groups}
\tablewidth{0pt} 
\tablehead{ 
\colhead{Name} &
\colhead{$r_c$} &
\colhead{$\beta$} &
\colhead{$S_0$}  \\
\colhead{} &
\colhead{(kpc)} &
\colhead{} &
\colhead{counts/kpc} 
}
\startdata
XBS 02&      35.07$\pm$4.75&    0.61$\pm$0.08&   0.0027$\pm$0.0004\\
XBS 06&      143.28$\pm$3.22&   0.76$\pm$0.02&   0.0255$\pm$0.0006\\
XBS 09&      57.40$\pm$28.36&    0.51$\pm$0.25&   0.0008$\pm$0.0004\\
XBS 11&      67.58$\pm$9.80&    0.48$\pm$0.07&   0.0087$\pm$0.0013\\
XBS 13&      22.05$\pm$1.79&    0.43$\pm$0.03&   0.0366$\pm$0.0030\\
XBS 17&      89.40$\pm$14.39&    0.40$\pm$0.06&   0.0010$\pm$0.0002\\
XBS 18&      28.38$\pm$18.87&    0.48$\pm$0.32&   0.0873$\pm$0.0581\\
XBS 20&      64.20$\pm$11.24&    0.59$\pm$0.10&   0.0106$\pm$0.0019\\
XBS 27&      95.37$\pm$44.82&    0.94$\pm$0.44&   0.0161$\pm$0.0076\\
XBS 41&      109.75$\pm$85.61&   0.48$\pm$0.37&   0.0039$\pm$0.0030\\
XBS 42&      40.11$\pm$2.85&    0.81$\pm$0.06&   0.0061$\pm$0.0004\\ 
 \enddata
\label{tab_bp}
\end{deluxetable}

\begin{figure}[!htb]
\begin{center}
\includegraphics[width=74mm]{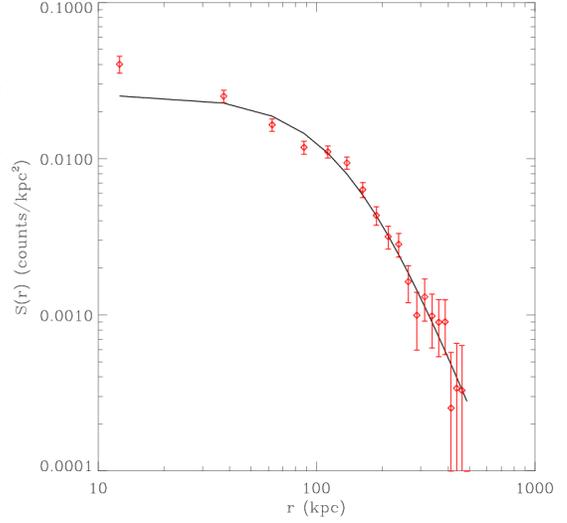}
\end{center}
\caption{Surface brightness profile of the brightest group XBS 06.  The $\beta$-Model (black line) is characterized by $r_c = 143.3\pm3.22$ kpc, $\beta = 0.76\pm0.02$ and $S_0 = 0.03\pm0.0006$ counts/kpc$^2$.}
\label{surf_bri}
\end{figure}

\par To determine the 0.5-2 keV X-ray luminosity for the 31 groups with determined redshifts, we follow the procedure described in \citet{2000Bohringer}.  The analysis of the X-ray emission for each group consists basically of measuring the net source  counts in a specified region of the galaxy group and then converting the count rate into X-ray flux and luminosity.

\par The first step is to estimate the background counts used to obtain the net source counts for each group.  The background contribution in the 0.5-2 keV band is evaluated in an annulus centered on the central coordinates of the group, with inner and outer radii equal to 0.5 Mpc $+$ 1 arcminute and 0.5 Mpc $+$ 2 arcminute\footnote{For sources XBS 06, XBS 17, we took the inner and outer radii equal to 1 Mpc $+$ 1 arcminute and 1 Mpc $+$ 2 arcminute, respectively, because their $R_{500}$ is larger than 0.5 Mpc.}, respectively. The annulus is divided into 12 equal sectors.  We obtain the count rate of each sector by dividing the photon counts by the average exposure time of each sector.  We excluded all counts coming from the point sources catalogued in \citet{2005Kenter} from both the background and source counts.  In addition, we visually inspected the deep fields in the Bo\"otes region to eliminate the point sources that were not catalogued by \citet{2005Kenter}.  To avoid the survey border effect and the contamination by other extended sources inside the ring, the median sector count rate was determined and sectors with lower than 2$\sigma$ or higher than 3$\sigma$ deviation from the median of all the sectors were discarded from further calculations.  The final background count rate is given by the median count rates of the remaining sectors.

\par The X-ray count rates of the galaxy groups are estimated inside 0.5 Mpc as well as $R_{500}$, when possible. The background count rate normalized by the source area is subtracted from the source count rate.  To convert the net count rates to X-ray flux in the 0.5-2 keV energy band, we use the \texttt{PIMMS}\footnote{Available on the HEASARC-NASA website} software package routine. We calculate the conversion factor from count rates to flux for a source assuming a given spectral model, temperature, abundance and hydrogen column density.  We adopted the astrophysical plasma emission code APEC \citep{2001Smith} to represent the intra-group medium, with a metalicity equal to 0.3$Z_\odot$ and  a temperature of 1 keV.  The hydrogen column density (21 cm) in the direction of each group was obtained from \citet{1990Dickey}.  Once the redshift of the group is known, we determine the luminosity, L$_{X1}$, based on the measured flux. We use this luminosity to better estimate the temperature of the gas using the L$_X$-T$_X$ relation from \citet{2012Sun}:

\begin{equation}
E(z)^{-1} L_X = L_0 \left( \frac{T_X}{2.5keV} \right) ^{2.74}
\end{equation}

\noindent where T$_X$ is in keV, L$_o$ is equal to 0.3334$\times$ 10$^{44}$ ergs s$^{-1}$ and $E(z) = \sqrt{\Omega_m (1 + z)^3 + \Omega_\Lambda}$.  The temperature estimate allows the calculation of a new flux and new luminosity.  We repeat this procedure iteratively, until the difference between the temperatures of the two last iterations is less than 0.1 keV. Usually, the temperature converges in two or three iterations. The $k$-correction from \citet{2000Bohringer} is applied to obtain the rest-frame X-ray luminosity.  To convert the X-ray luminosity in the 0.5-2 keV energy band into bolometric luminosity, we used \texttt{xspec} to simulate the spectrum of each group with the appropriate nH column density, redshift, temperature and abundance. We assumed a thermal plasma model and a photoelectric absorption model for the intragroup medium. We calculated the ratio between the bolometric (0.1-100 keV) and 0.5-2 keV energy band luminosity and multiplied the rest-frame X-ray luminosity of the group by this ratio, to convert it to bolometric luminosity.  The results are shown in Table \ref{tab_xray}.

\par We also extract the X-ray surface brightness profiles for the groups.  Using \texttt{SHERPA}\footnote{http://cxc.harvard.edu/sherpa4.4/index.html}, we fit a $\beta$-model \citep{1976Cavaliere} to the surface brightness profile:

\begin{equation}
S(r) = S_0 \left( 1 + \frac{r^2}{r_{c}^{2}}\right) ^{-3\beta + \frac{1}{2}}
\end{equation}

\noindent where $r_c$ is the core radius and $S_o$ is the central surface brightness.  The background was estimated using an annulus, with inner and outer radii equal to 0.5 Mpc $+$ 1 arcminute and 0.5 Mpc $+$ 2 arcminute\footnote{For sources XBS 06, XBS 17, the background was estimated inside a annulus, with inner and outer radii equal to 1 Mpc $+$ 1arcminute and 1 Mpc $+$ 2 arcminute}.  The background region of each group was chosen to be free from contamination of extended sources and far from the detector edge.  The counts from point sources were excluded from both the background and source regions.  Figure \ref{surf_bri} shows the $\beta$-model fit for the group XBS 006.  We were able to extract the parameters of the $\beta$-model from the brightness profiles for 11 of 30 groups, which have at least 200 counts.  For groups with counts below 200 photons, it was not possible to determine these parameters.  The results are shown in Table \ref{tab_bp}.  We find that 11 groups have $\beta$ slopes between 0.4 and 0.9, while the core radii range from 22 to 143 kpc.

\par Of these 11 groups, we could determine the gas temperature for 9.   We use CIAO to extract the spectrum of each group inside 0.5 Mpc and $R_{500}$, using the blank sky fields to subtract the background.  The background region has the same area as its respective source region.  The procedure differs from our X-ray luminosity and X-ray surface brightness profile analysis, since now it is necessary to use the blank sky to subtract the background, because most of data used to extract the spectra are the smaller fields from ACIS-S observations.   After extracting the spectrum, we used \texttt{XSPEC} to fit a thermal model and obtain the temperature of the gas.  As we did to convert X-ray luminosity in the 0.5-2 keV energy band into bolometric luminosity, we assume a thermal plasma and photoelectric absorption model for the intra-group medium.  We use the nH column density and redshift of each group.  As a first trial, we chose T$_X$ to be 1 keV and the heavy element abundance to be 0.3Z$_\odot$ and then constrained those parameters by the spectral fit.  We were able to fit the spectrum of 9 groups.  The gas temperatures of these groups range from 0.6 to 4 keV.  The results are shown in Table \ref{tab_xray}.  

\begin{deluxetable}{lcc}
\tablecaption{X-ray Properties Inside 0.5 Mpc}
\tablewidth{0pt} 
\tablehead{ 
\colhead{Name} &
\colhead{$L_X^{bol}$} &
\colhead{$T_X$}  \\
\colhead{} &
\colhead{(10$^{42}$ ergs s$^{-1}$)} &
\colhead{(keV)} 
}
\startdata
XBS02&        14.85  $\pm$ 2.13    &     0.82$\pm$1.38 \\
XBS04&        156.00$\pm$ 70.20     &     ---- \\
XBS05&        14.00  $\pm$ 4.85    &     ---- \\
XBS06&        30.65 $\pm$ 1.12    &     3.53$\pm$0.69 \\
XBS07&        0.48   $\pm$ 0.89    &     0.66$\pm$0.05 \\
XBS08&        1.14   $\pm$ 0.17    &     ---- \\
XBS09&        6.38   $\pm$ 1.11    &     ---- \\
XBS11&        23.5 3 $\pm$ 11.64   &     1.99$\pm$0.55 \\
XBS13&        71.18  $\pm$ 0.51    &     2.18$\pm$0.54 \\
XBS14&        49.17  $\pm$ 9.53    &     ---- \\
XBS17&        13.86  $\pm$ 4.02    &     ---- \\
XBS18&        22.20  $\pm$ 1.22    &     3.37$\pm$0.23 \\
XBS20&        16.40  $\pm$ 8.25    &     1.36$\pm$0.20 \\
XBS22&        39.91  $\pm$ 7.74    &     ---- \\
XBS25&        11.75  $\pm$ 2.78    &     ---- \\
XBS26&        1.99   $\pm$ 0.62    &     ---- \\
XBS27&        106.10  $\pm$ 9.5    &     3.63$\pm$0.74 \\
XBS28&        111.07 $\pm$ 12.91   &     ---- \\
XBS29&        8.76   $\pm$ 26.60   &     ---- \\
XBS32&        113.87 $\pm$ 30.51   &     ---- \\
XBS33&        52.73  $\pm$ 12.75   &     ---- \\
XBS35&        0.53   $\pm$ 0.08    &     ---- \\
XBS36&        4.40   $\pm$ 6.82    &     ---- \\
XBS37&        7.97   $\pm$ 1.29    &     ---- \\
XBS38&        48.90  $\pm$ 11.39   &     ---- \\
XBS39&        24.96  $\pm$ 6.45    &     ---- \\
XBS41&        25.47  $\pm$ 4.07    &     4.02$\pm$1.08 \\
XBS42&        21.03  $\pm$ 4.66    &     ---- \\
XBS43&        82.50  $\pm$ 17.24   &     ---- \\
XBS46&        6.15   $\pm$ 1.99    &     ---- \\
XBS52&        35.09  $\pm$ 26.02    &     ---- \\
 \enddata
\label{tab_xray}
\end{deluxetable}

%%%%%%%%%%%%%%%%%%%%%%%%%%%%%%%%%%%%%%%%%%%%%%%%%%%%%%%%%%%%%%%%%%%%%%%%
%%%%%%%%%%%%%%%%%%%%%%%%%%%%%%%%%%%%%%%%%%%%%%%%%%%%%%%%%%%%%%%%%%%%%%%%
%%%%%%%%%%%%%%%%%%%%%%%%%%%%%%%%%%%%%%%%%%%%%%%%%%%%%%%%%%%%%%%%%%%%%%%%
%%%                                                                  %%%
%%%                         Scaling Relations                        %%%
%%%                                                                  %%%
%%%%%%%%%%%%%%%%%%%%%%%%%%%%%%%%%%%%%%%%%%%%%%%%%%%%%%%%%%%%%%%%%%%%%%%%
%%%%%%%%%%%%%%%%%%%%%%%%%%%%%%%%%%%%%%%%%%%%%%%%%%%%%%%%%%%%%%%%%%%%%%%%
%%%%%%%%%%%%%%%%%%%%%%%%%%%%%%%%%%%%%%%%%%%%%%%%%%%%%%%%%%%%%%%%%%%%%%%% 

\section{Scaling Relations}

\begin{deluxetable*}{llccc}
\tablecaption{Scaling Relations of Luminosity, Velocity Dispersion and Mass}
\tablewidth{0pt} 
\tablehead{ 
\colhead{Relation (X-Y)} &
\colhead{Sample} &
\colhead{A} &
\colhead{B} &
\colhead{$\sigma_{LogY|X}$}
}
\startdata
L$_{opt}^{R500}$--$\sigma^{R500}$& XBS        & 2.79$\pm$0.05& 0.79$\pm$0.09& 0.15\\
                                & NoSOCS     & 2.79$\pm$0.01& 0.47$\pm$0.04& 0.22\\
                                & XBS, NoSOCS& 2.79$\pm$0.01& 0.55$\pm$0.04& 0.21\\
                                & XBS, NoSOCS (M$_{500} >$ 5$\times$10$^{13}$ M$_{\odot}$)& 2.79$\pm$0.01& 0.42$\pm$0.03& 0.21\\
                                & XBS, NoSOCS (M$_{500} \leq$ 5$\times$10$^{13}$ M$_{\odot}$)& 2.55$\pm$0.24& 0.43$\pm$0.36& 0.47\\
                                &                                                   &              &              &              \\
L$_{opt}^{R500}$--M$_{R500}$& XBS        & 0.15$\pm$0.12& 1.65$\pm$0.17& 0.23\\
                           & NoSOCS     & 0.52$\pm$0.03& 1.37$\pm$0.13& 0.19\\
                           & XBS, NoSOCS& 0.53$\pm$0.04& 1.65$\pm$0.12& 0.20\\
                           & XBS, NoSOCS (M$_{500} >$ 5$\times$10$^{13}$ M$_{\odot}$)& 0.51$\pm$0.03& 1.13$\pm$0.08& 0.19\\
                           & XBS, NoSOCS (M$_{500} \leq$ 5$\times$10$^{13}$ M$_{\odot}$)& 0.11$\pm$0.37& 1.50$\pm$0.50& 0.29\\
                                &                                                   &              &              &              \\
L$_{X}^{R500}$--$\sigma^{R500}$& XBS        & 2.16$\pm$0.07& 0.23$\pm$0.07& 0.30\\
                              & NoSOCS     & 2.36$\pm$0.06& 0.20$\pm$0.03& 0.49\\
                              & XBS, NoSOCS& 2.27$\pm$0.05& 0.24$\pm$0.03& 0.57\\
                              & XBS, NoSOCS (M$_{500} >$ 5$\times$10$^{13}$ M$_{\odot}$)& 2.47$\pm$0.04& 0.14$\pm$0.02& 1.01\\
                              & XBS, NoSOCS (M$_{500} \leq$ 5$\times$10$^{13}$ M$_{\odot}$)& 2.19$\pm$0.05& 0.10$\pm$0.05& 1.38\\
                                &                                                   &              &              &              \\
L$_{X}^{R500}$--M$_{R500}$& XBS        & -1.25$\pm$0.21& 0.66$\pm$0.21& 0.34\\
                         & All        & -1.05$\pm$0.05& 0.68$\pm$0.02& 0.74\\
                         & All (M$_{500} >$ 5$\times$10$^{13}$ M$_{\odot}$)& -0.85$\pm$0.05& 0.61$\pm$0.02& 0.74\\
                         & All (M$_{500} \leq$ 5$\times$10$^{13}$ M$_{\odot}$)& -1.19$\pm$0.08& 0.45$\pm$0.08& 0.76\\
 \enddata
\label{tab_scaling}
\end{deluxetable*}

\begin{figure*}%[!htb]
\begin{center}
\includegraphics[width=98mm]{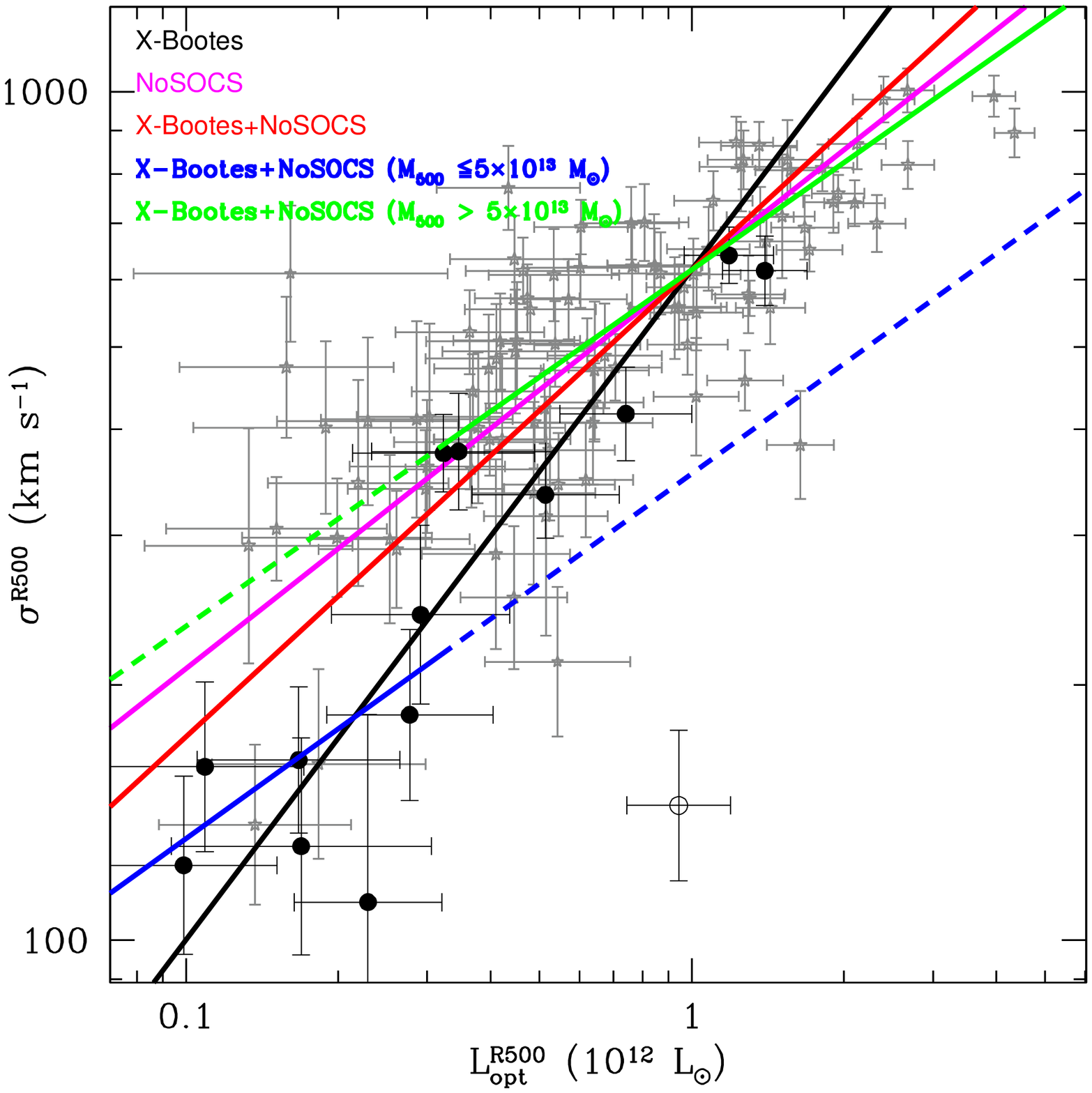}
\includegraphics[width=98mm]{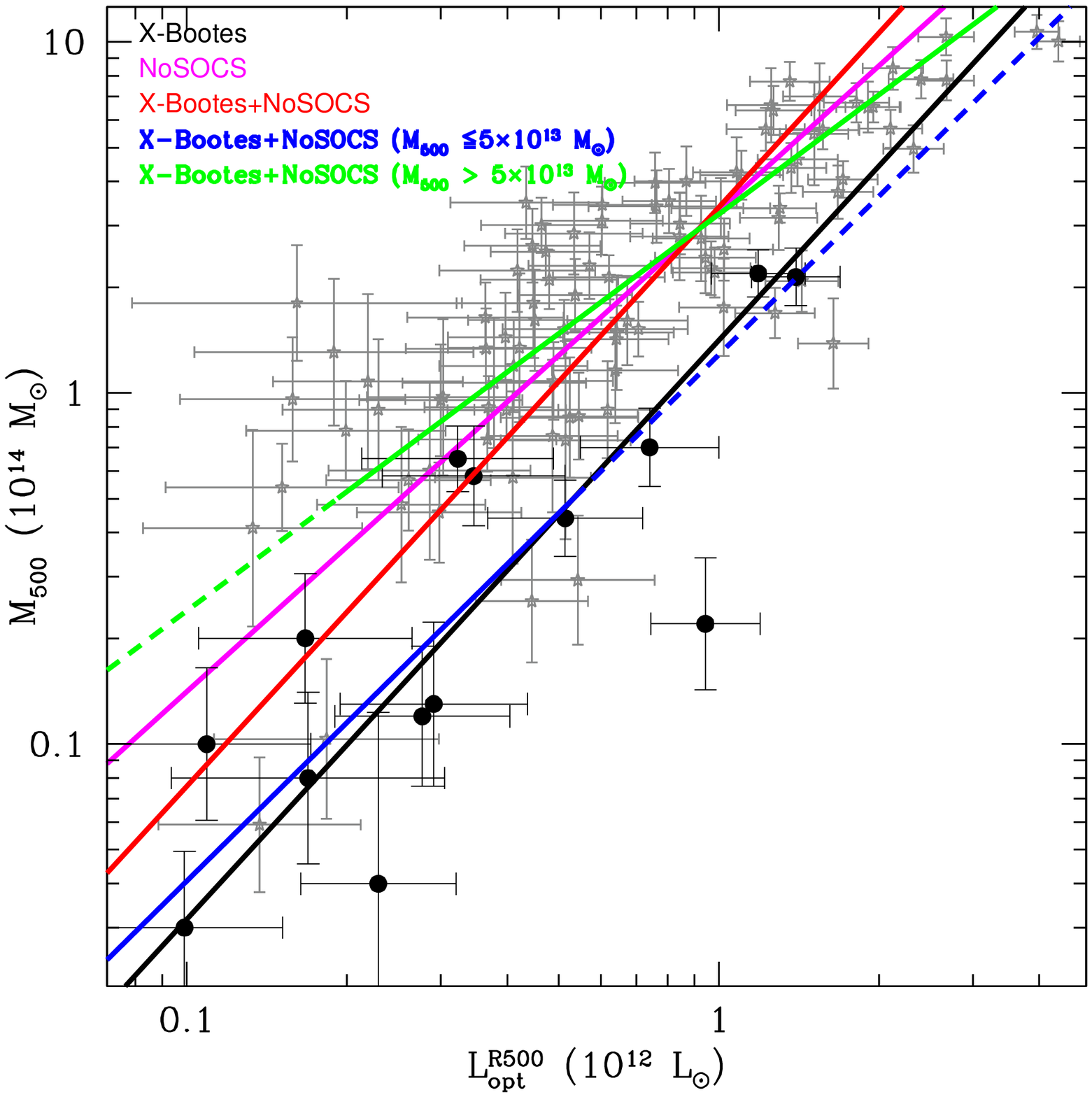}
\end{center}
\caption{Comparison between optical properties for 14 X-Bo\"otes groups at z $<$ 0.35, with at least five member galaxies (black dots) and the NoSOCS sample of galaxy groups and clusters (gray stars). In the upper panel, we show the relation between optical luminosity and velocity dispersion ($\sigma$-L$_{opt}$).  In the lower panel, we show the relation between optical luminosity and mass (M$_{500}$-L$_{opt}$).  The black line is the orthogonal regression fit using only the X-Bo\"otes data.  The magenta line is the regression fit only for NoSOCS. The red, blue and green lines are, respectively, the orthogonal regression fit for both data sets (X-Bo\"otes and NoSOCS), both data sets with M$_{500} \leq$ 5$\times$10$^{13}$ M$_{\odot}$ and both data sets with M$_{500} >$ 5$\times$10$^{13}$ M$_{\odot}$.}
\label{opt_rel}
\end{figure*}

\begin{figure*}%[!htb]
\begin{center}
\includegraphics[width=98mm]{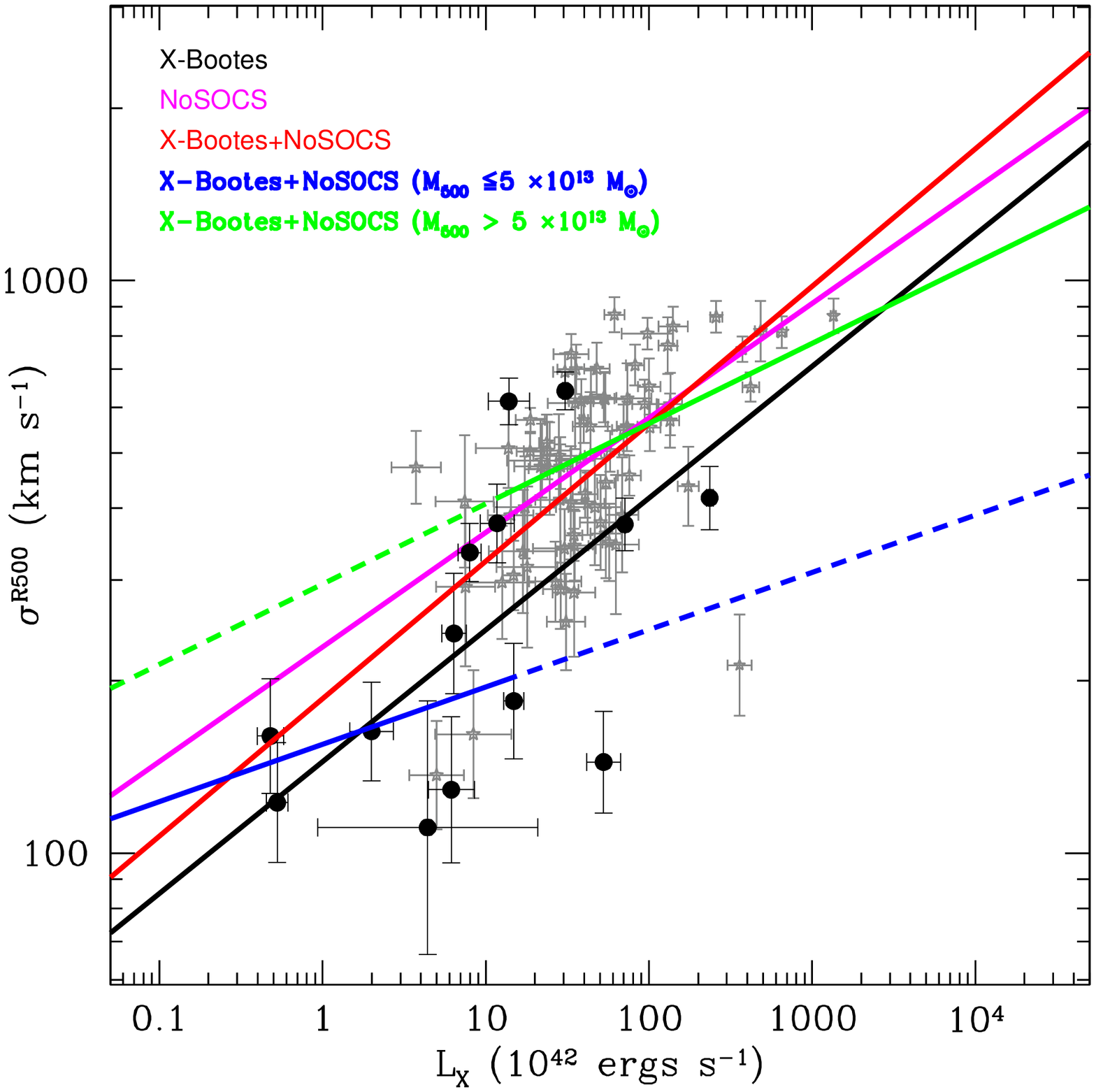}
\includegraphics[width=98mm]{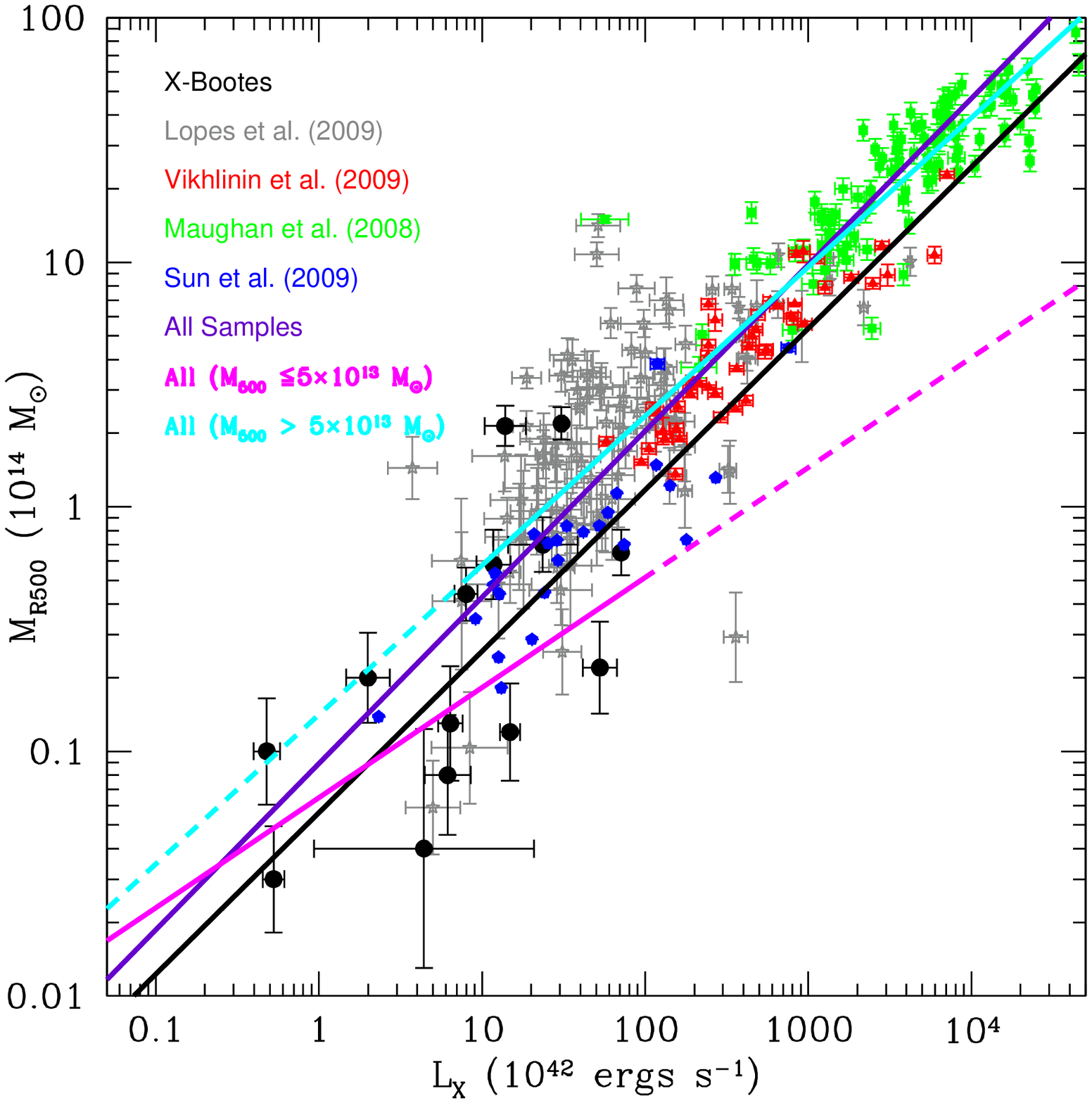}
\end{center}
\caption{Comparison between X-ray properties for 14 X-Bo\"otes groups at z $<$ 0.35, with at least five member galaxies (black dots) and the NoSOCS sample (gray stars). In the upper panel, we show the relation between bolometric X-ray luminosity and velocity dispersion ($\sigma$-L$_{X}$).  In the lower panel, we show the relation between bolometric X-ray luminosity and total mass derived from the velocity dispersion (M$_{500}$-L$_{X}$) and we also include he results from \citet{2009Vikhlinin} (red triangles), \citet{2008Maughan} (green squares) and \citet{2009Sun} (blue pentagons).  The black line is the orthogonal regression fit using only the X-Bo\"otes data. The purple, magenta and cyan lines are, respectively, the orthogonal regression fit for all data, all data with M$_{500} \leq$ 5$\times$10$^{13}$ M$_{\odot}$ and all data with M$_{500} >$ 5$\times$10$^{13}$ M$_{\odot}$.}
\label{xray_rel}
\end{figure*}

\par In this section, we explore the use of optical and X-ray properties as mass proxies at $z < 0.35$.  We compare $M_{500}$ and $\sigma_{gr}$ to optical ($L_{opt}$) and X-ray ($L_X$) luminosities by fitting an orthogonal regression \citep{1996Akritas}.  All the scaling relations obtained in this work are of the form:

\begin{equation}
{\rm Log_{10}}(Y) = A + B\times {\rm Log_{10}}(X).
\end{equation}

\noindent The result of the scaling relations are summarized in Table \ref{tab_scaling}.  The columns give (1) the parameters in the relation (\textit{X-Y}); (2) Sample(s) used; (3) and (4) the intercept (\textit{A}) and the slope (\textit{B}) and their respective uncertainties and (5) the scatter in the \textit{Y} parameter at a fixed \textit(X).

\par Figure \ref{opt_rel} shows the comparison between $\sigma_{gr}$ and $L_{opt}$ (upper panel) and between $M_{500}$ derived from the velocity dispersion and $L_{opt}$ (lower panel).  All properties were computed within $R_{500}$.  We also compare the X-Bo\"otes sample (black dots) with the NoSOCS sample (gray stars) from \citet{2009aLopes}.  To make the NoSOCS sample comparable to the X-Bo\"otes sample, the velocity dispersions were recomputed inside $R_{500}$.

\par We fit the relations for the Bo\"otes sample (black line), for NoSOCS sample (magenta line), Bo\"otes and NoSOCS with no cut in mass (red line), with a mass cut of $M_{500} \leq 5\times10^{13} M_{\odot}$ (blue line) and with a mass cut of $M_{500} > 5\times10^{13} M_{\odot}$ (green line).  The best fit values of the relations $L_{opt}-\sigma_{gr}$ and $L_{opt}-M_{500}$ for the Bo\"otes sample do not agree with the results for NoSOCS.  While the slope of the $L_{opt}-\sigma_{gr}$ relation for Bo\"otes has a slope of 0.79$\pm$0.09, the relation for NoSOCS has a slope of 0.47$\pm$0.04.  There is also a large scatter in the relation for $\sigma < 400$ km s$^{-1}$.  This can be explained by the fact that low mass systems have fewer than 10 galaxy members selected and the use of systems with fewer than 10 galaxies leads to an increase in the scatter in the scaling relations.  Due to the large scatter and the few points below $M_{500} \leq 5\times10^{13}$ M$_{\odot}$, it is not possible to say if there is a break in the $L_{opt}-\sigma_{gr}$ relation between the high and low-mass systems.

\par Optical scaling relations are generally more difficult to interpret, because their behavior cannot be described by simple physics scaling arguments.  This comes from the fact that galaxy properties are the result of a complex non-linear process of formation and evolution.  According to the self-similar model prediction, the $M_{Tot}$-$L_{opt}$ relation has a predicted power law slope equal to one.  For this to be true, the mass-to-light ratio must be constant for the sample. As most studies indicate (e.g. \citet{2009bLopes}), there is an increase of $M/L_{opt}$ with the mass of the cluster.  The straightforward result of the dependence of $M/L_{opt}$ on the mass is that the power law slope of the $M_{Tot}$-$L_{opt}$ relation will not be the same for galaxy groups and clusters.

\par In Figure \ref{xray_rel}, we show the $L_{X}-\sigma_{gr}$ (upper panel) and $L_{X}-M_{500}$ (lower panel) relations.  All properties were computed within $R_{500}$.  We also compare the X-Bo\"otes sample (black dots) with the samples from \citet{2009aLopes} (gray stars), \citet{2009Vikhlinin} (red triangles), \citet{2008Maughan} (green squares) and \citet{2009Sun} (blue pentagons).  We fit the relations for the Bo\"otes sample (black line), for all the samples (purple line)and for all the samples with mass cuts of $M_{500} \leq 5\times10^{13} M_{\odot}$ (magenta line) and $M_{500} > 5\times10^{13} M_{\odot}$ (cyan line). It is important to note that differently from X-Bo\"otes and NoSOCS samples, where masses were determined using galaxy dynamics, \citet{2009Vikhlinin, 2009Sun, 2008Maughan} estimated the $M_{500}$ using X-ray data. To verify how the results change if the X-ray samples are neglected, we fit the $L_{X}-M_{500}$ relation using only the X-Bo\"otes and NoSOCS samples. The best fit values are $M_{500} \propto L_{X}^{0.76\pm0.07}$ (for no mass cut), $M_{500} \propto L_{X}^{0.37\pm0.13}$ (for $M_{500} \leq 5\times10^{13}$ M$_\odot$) and $M_{500} \propto L_{X}^{0.50\pm0.05}$ (for $M_{500} > 5\times10^{13}$ M$_\odot$). These values are consistent within $1\sigma$ with the results displayed in Table \ref{tab_scaling}.

\par Since the X-Bo\"otes sample was X-ray selected, it is clearly not volume limited, but flux limited and therefore various correlations could be subject to Malmquist bias.  To test the Malmquist bias in the $L_{X}-\sigma_{gr}$ and $L_{X}-M_{500}$ relations, we choose two volumes ($z = 0.20$ and $z = 0.35$) and determine the limiting X-ray luminosity ($L_{X} = 1.12\times 10^{42}$ and $L_{X} = 4.01\times 10^{42}$ ergs s$^{-1}$, respectively) equivalent to the limiting flux of the X-Bo\"otes survey for extended sources ($f_X =1\times10^{-14}$ ergs s$^{-1}$ cm$^{-2}$) in each volume.  For $z \leq 0.35$ and $L_{X} > 4.01\times 10^{42}$ ergs s$^{-1}$, the best fit values\footnote{In this fit, the groups XBS 07, XBS 26 and XBS 35 were excluded} are $\sigma_{gr} \propto L_{X}^{0.16 \pm 0.12}$ and $M_{500} \propto L_{X}^{0.83 \pm 0.59}$.  And for $z \leq 0.20$ and $L_{X} > 1.12\times 10^{42}$ ergs s$^{-1}$, the best fit values\footnote{In this fit, the groups XBS 02, XBS 07, XBS 11, XBS 33, XBS 35 and XBS 36  were excluded} are $\sigma_{gr} \propto L_{X}^{0.42 \pm 0.27}$ and $M_{500} \propto L_{X}^{1.13 \pm 0.65}$.  The resulting power law exponents of the scaling relations are consistent with the best fit values found for the X-B\"ootes sample (see Table \ref{tab_scaling}).  The uncertainties on the fits are large and these are directly linked to the small number of sources used to fit the relations. For this reason, we cannot conclude that we are not entirely free from the Malmquist bias, but the tests suggest that we are not dominated by it. To investigate it in details, it is necessary more data and larger samples to test this further.

\par The best fit values of the relation $L_{X}-\sigma_{gr}$ for the Bo\"otes sample agree with the results for all the samples (NoSOCS and X-Bo\"otes combined), with the self-similar model predictions ($\sigma \propto L_{X}^{1/4}$) and with the results found in Figure 8 of \citet{2008Fassnacht}, in \citet{2001Mahdavi} ($\sigma_{gr} \propto L_{X}^{0.23_{-0.03}^{+0.02}}$) and in \citet{2000Helsdon} ($\sigma_{gr} \propto L_{X}^{0.21\pm0.03}$).  The slope of the $L_{X}-\sigma_{gr}$ relation for Bo\"otes has a slope of 0.23$\pm$0.07.  When we considered the mass cut $M_{500} = 5\times10^{13}$ M$_{\odot}$, the result is a similar slope for both high and low mass regimes (0.14 and 0.10, respectively).  For the high mass systems, the slope may have been flattened by 6 clusters from the NoSOCS sample.  They have high values for $L_X$ (10$^{44}$-10$^{45}$ ergs s$^{-1}$) compared to their velocity dispersions (700-1000 km s$^{-1}$). When these 6 clusters are excluded from the fit, the relation has a slope of 0.25$\pm$0.05.  In the low mass regime ($M_{500} \leq 5\times10^{13}$ M$_{\odot}$), the slope of the relation is obtained using 16 groups and therefore the result is uncertain.  So, based on these results, we can not conclude if there is a break in the $L_{X}-\sigma_{gr}$ relation.

\par Although we are unable to determine the existence of a break in the $L_{X}-\sigma_{gr}$ relation, there are a set of galaxy groups with $\sigma_{gr} \leq 200$ km s$^{-1}$ that has an offset in an opposite direction to what would be expected if we consider the effects of feedback processes over the intragroup medium.  In the presence of AGN outburst, for instance, a fraction of the intragroup gas could be expelled from the group gravitational potential.  As a direct consequence, the group $L_X$ would decrease, but the velocity dispersion would remain unaltered.  But, what is observed from the upper panel of Figure \ref{xray_rel} is that groups with $\sigma_{gr} \leq 200$ km s$^{-1}$ have higher $L_X$ in comparison with their $\sigma_{gr}$.  This same phenomenon was observed before by \citet{1994dellAntonio,2000Mahdavi,2001Mahdavi,2005Helsdon,2010Rines}.

\par If the velocity dispersion values estimated from the galaxy velocities in the line of sight (see $\S$3.2) are a fair estimate of the 'real' velocity dispersion, then some process must have reduced the velocity dispersion of the galaxies in these systems.  \citet{2005Helsdon} suggested three scenarios to explain the reduction of the galaxy velocity dispersion.  The first possibility is the dynamical friction, which leads to a transfer of energy from a large orbiting body to the dark matter particles through which it moves.  The second possibility is that orbital energy may be converted into internal energy of galaxies, through tidal interactions.  This effect is not significant in clusters, because the orbital velocities of galaxies are substantially higher than their internal velocity dispersion.  The third possibility is that much of the orbital motion of the galaxy groups takes place in the plane of the sky and does not contribute to the line of sight velocity dispersion.  \citet{2002Tovmassian} showed that it is expected to find elongation and an anisotropic velocity dispersion tensor in many systems, since groups generally form within cosmic filaments.

\par For the $L_{X}-M_{500}$ relation (bottom panel of Figure \ref{xray_rel}), the best fit values for the X-Bo\"otes sample agree with the results found for all the samples and for all the samples with the mass cut of $M_{500} > 5\times10^{13}$ M$_{\odot}$.  However, in the low mass regime ($M_{500} \leq 5\times10^{13}$ M$_{\odot}$), the slope of 0.45$\pm$0.08 does not agree with the steeper slopes (0.61-0.68) found for the other samples.  It is clear in Figure \ref{xray_rel} that there is a break in the $L_X-M_{500}$ relation between low and high mass systems.  Thus, galaxy groups cannot be described by the same power law as galaxy clusters.

\par A large number of researchers (e.g., \citet{2002Reiprich}; \citet{2002Ettori}; \citet{2006Maughan}; \citet{2007Maughan}; \citet{2009Vikhlinin}; \citet{2009Pratt}) have measured the slope for $L_{X}-M_{Tot}$ relation as 0.5-0.7, which is flatter than the self-similar predictions of $M_{Tot} \propto L_{X}^{3/4}$. These works show that the ICM is heavily affected by non-gravitational processes. Since $L_X$ comes from the hot ICM, any change in the amount of hot gas in the system will influence the $L_X$.  But, the feedback processes cannot explain the steep $L_{X}-M_{500}$ relation for low mass systems.  If we consider the $L_{X}-M_{500}$ relation for clusters (purple and cyan lines in Figure \ref{xray_rel}), the groups show an excess of $L_X$ compared to $M_{500}$.  Since the $M_{500}$ of the X-B\"ootes groups were derived from the velocity dispersion (see $\S$3.2), reduction in the velocity dispersion will provoke an reduction in $M_{500}$.  As we mentioned before, dynamical friction, tidal interactions between the galaxy members of the group and projection effects can decrease the velocity dispersion values. These effects are not significant in clusters, causing the $L_{X}-M_{500}$ to be flatter for groups than for clusters.

\par It is interesting to note that the steepening of scaling relations in the group regime extends also to systems at intermediate and high redshifts. \citet{2005Willis} found a steepening of the $L_X-T_X$ relation for poor groups at $z\sim0.4$. \citet{2007Jeltema, 2009Jeltema} showed that there is some tendency for intermediate-redshift groups to have velocity dispersions that are low given their X-ray luminosity. \citet{2011Adami} found signs of major substructures in the velocity distribution of high-redshift groups which could explain their relatively high $L_X$ value compared to its optical richness. All these works show some indications of the deviation from the energy equipartition between galaxies and intragroup gas.

%%%%%%%%%%%%%%%%%%%%%%%%%%%%%%%%%%%%%%%%%%%%%%%%%%%%%%%%%%%%%%%%%%%%%%%%
%%%%%%%%%%%%%%%%%%%%%%%%%%%%%%%%%%%%%%%%%%%%%%%%%%%%%%%%%%%%%%%%%%%%%%%%
%%%%%%%%%%%%%%%%%%%%%%%%%%%%%%%%%%%%%%%%%%%%%%%%%%%%%%%%%%%%%%%%%%%%%%%%
%%%                                                                  %%%
%%%                           Log N- Log S                           %%%
%%%                                                                  %%%
%%%%%%%%%%%%%%%%%%%%%%%%%%%%%%%%%%%%%%%%%%%%%%%%%%%%%%%%%%%%%%%%%%%%%%%%
%%%%%%%%%%%%%%%%%%%%%%%%%%%%%%%%%%%%%%%%%%%%%%%%%%%%%%%%%%%%%%%%%%%%%%%%
%%%%%%%%%%%%%%%%%%%%%%%%%%%%%%%%%%%%%%%%%%%%%%%%%%%%%%%%%%%%%%%%%%%%%%%% 

\section{Log N- Log S and The X-ray Luminosity Function of Groups}
\label{logn-lgos}

\begin{figure}
\begin{center}
\includegraphics[width=89mm]{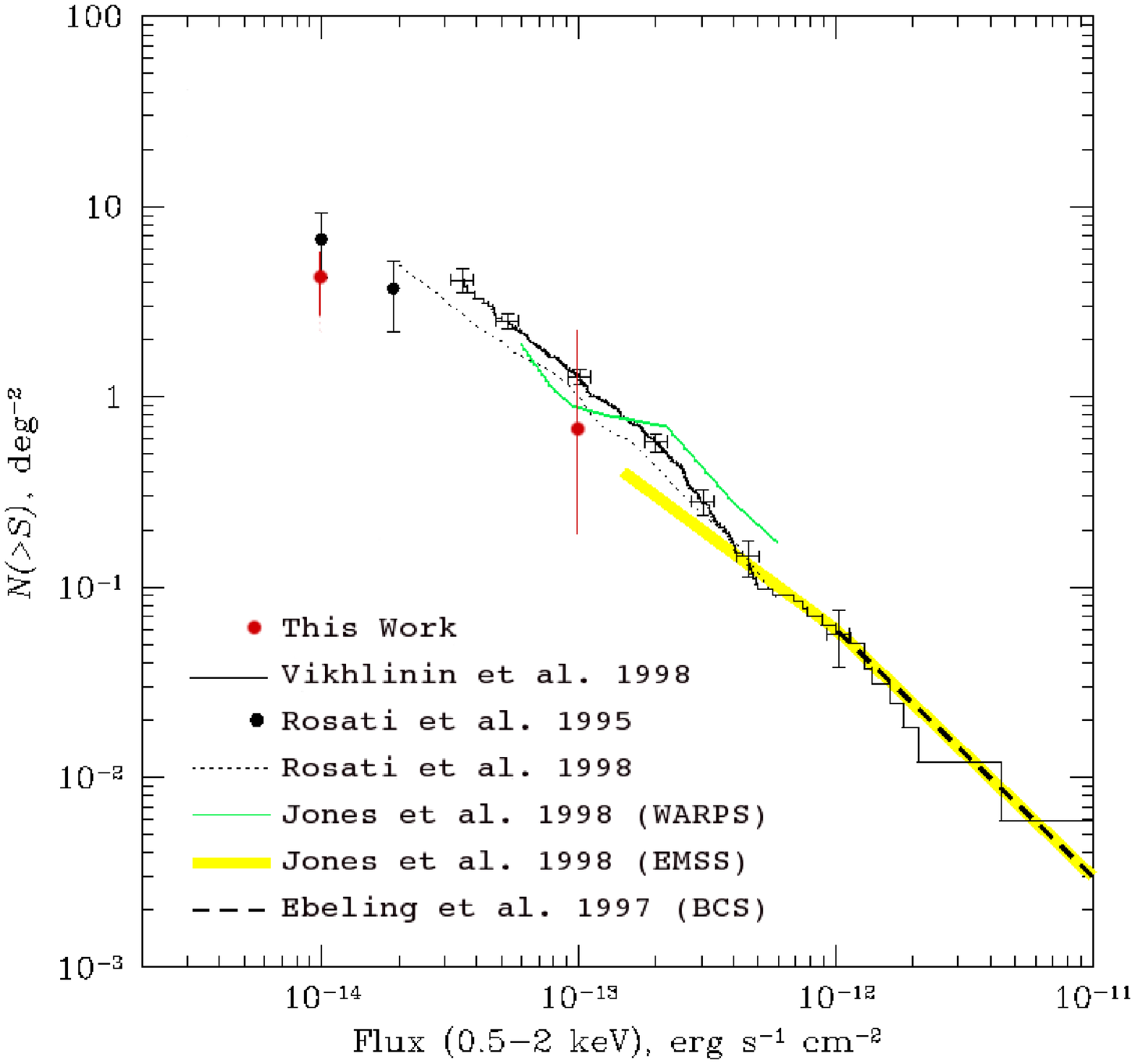}
\end{center}
\caption{Log \textit{N}-Log \textit{S} relation. The results from X-Bo\"otes survey are shown as the red dots including error
bars. Vertical error bars represent the uncertainty in the number of groups. We compare our results with other surveys: \citet{1998Vikhlinin} (black solid histogram with several individual points including error bars); \citet{1995Rosati} (black dot); \citet{1998Rosati} (black short dashed line); WARPS (green solid line); EMSS (yellow heavy solid line) and BCS (black long dashed line).}
\label{logn-logs_plot}
\end{figure}

\begin{figure}
\begin{center}
\includegraphics[width=90mm]{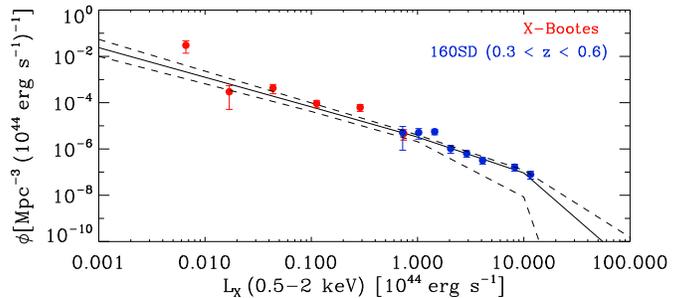}
\end{center}
\caption{Determinations of the group/cluster X-ray luminosity function measured by X-Bo\"otes (red circles) and 160SD (blue circles).  The Schechter fit for the nonparametric data points is represented by the black line. The dashed lines represent the $1\sigma$ uncertainty region of the Schechter fit, assuming the errors for $L_X^*$ and $\alpha$ are correlated. The data point uncertainties are $\pm 1\sigma$.}
\label{XLF_fig}
\end{figure}

\begin{deluxetable*}{lcccc}
\tablewidth{0pt}
\tablecolumns{4}
\tablecaption{Expected Number of Groups Between $L_{X}^{min} < L_{X} < L_{X}^{max}$ Inside a Volume ($z_{min} < z < z_{max}$)\label{tab_nexp}}
\tablehead{
\colhead{$L_{X}^{min}$ - $L_{X}^{max}$} &
\multicolumn{3}{c}{$N_{exp}$} \\
\colhead{} &
\colhead{} &
\colhead{} &
\colhead{}&
\colhead{} \\
\colhead{(ergs s$^{-1}$)} &
\colhead{$0 < z \leq 0.046$} &
\colhead{$0.046 < z \leq 0.114$} &
\colhead{$0.114 < z \leq 0.238$} &
\colhead{$0.238 < z \leq 0.319$}
}
\startdata
\\
10$^{41}$ - 10$^{42}$        & 1178 &17    &       &   \\
10$^{42}$ - $5\times10^{42}$ & 474  & 6673 & 386   &   \\
$5\times10^{42}$ - 10$^{43}$ & 147  & 2067 & 16419 & 11
\enddata
\end{deluxetable*}

\par Using the survey solid angle, we calculate the number of groups per flux bin, the Log\textit{N}-Log\textit{S} relation, for the Bo\"otes group sample which is complete to $f_X =1\times10^{-14}$ ergs s$^{-1}$ cm$^{-2}$.  \citet{2005Kenter} estimated that the X-Bo\"otes survey on axis detection limit is $\approx 1\times 10^{-14}$ cgs (0.5-2 keV) for sources that are just demonstrably larger than the PSF. Since all the extended sources were detected at an existence significance threshold equivalent to $\approx 3\sigma$ and their Gaussian fitted profile have width broader than the local PSF\footnote{During the wavelet detection, \citet{2005Kenter} compared all detected sources with the local PSF. The profile of a detected source was fitted to a Gaussian with the width as a free parameter. The source was only considered extended, if the free parameter fitted width were larger than the local PSF width.}, our sample also is 100\% complete.  

\par The derived cumulative Log\textit{N}-Log\textit{S} plot is shown in Figure \ref{logn-logs_plot}. We also show the cluster counts derived in: \citet{1998Vikhlinin}, EMSS \citet{1998Jones}, the ROSAT All-Sky survey sample of X-ray brightest clusters (BCS; \citet{1997Ebeling}), the WARPS survey \citep{1998Jones} and the ROSAT cluster sample from Rosati et al. (1995, 1998).  Our cumulative numerical density of groups spans one order of magnitude in flux. At the faint end, our results are in excellent agreement with the samples of nearby clusters from Rosati et al. (1995,1998), \citet{1998Vikhlinin} and WARPS.

\par The Log\textit{N}-Log\textit{S} relation predicts the number of systems that future large X-ray surveys will observe. The eROSITA all sky survey, in particular, will have an average exposure time of 3 ks and will be complete to $f_X=3\times10^{-14}$ ergs s$^{-1}$ cm$^{-2}$. Thus, it will be able to observe clusters with masses as small as 3.5$\times$10$^{14}$ $h^{-1}$ M$_{\odot}$ to $z=2$.  Based on Figure \ref{logn-logs_plot}, eROSITA will detect 100,000-130,000  galaxy groups and clusters brighter than $f_X=3\times10^{-14}$ ergs s$^{-1}$ cm$^{-2}$ to $z=2$, what agrees with the predictions made by \citet{2012Chon} and \citet{2012Pillepich}.

\par With the eROSITA flux limit it will be possible to observe groups which are more luminous than $L_X$ = 10$^{41}$ ergs s$^{-1}$ to $z$ = 0.046, groups with $L_X$ = 10$^{42}$ ergs s$^{-1}$ to $z$ = 0.114, groups with $L_X$ = $5\times10^{42}$ ergs s$^{-1}$ to $z$ = 0.238 and groups with $L_X$ = 10$^{43}$ ergs s$^{-1}$ to $z$ = 0.319.  To estimate how these groups are distributed in bins of luminosity  within a given volume, we build the X-ray luminosity function (XLF).  

\par The differential luminosity function is defined as

\begin{equation}
\phi (L_X,z) = \frac{d^{2}N}{dVdL_X}(L_X,z),
\end{equation}

\noindent where $N$ is the number of clusters of luminosity $L_X$ in a volume $V$ at a redshift $z$. \citet{1968Schmidt} was the first to propose the $1/V_{max}$ technique for deriving a nonparametric representation of the differential cluster XLF. Later, this technique was generalized by \citet{1980Avni}. Here, we follow the procedure described in \citet{2004Mullis} to fit the XLF, using the X-Bo\"otes sample.  We divide the observed luminosity range into bins of width $\Delta L$, within each of which there are $N_j$ observed groups. The luminosity binning is uniform in log space. The XLF is estimated by summing the density contributions of each group in the given luminosity and redshift bin:

\begin{equation}
\phi (L_X, z) = \frac{1}{\Delta L}\sum^{N_j}_{i=1}\frac{1}{V_{max}(L_{X,i})}
\end{equation}

\noindent where $L_{X,i}$ is the luminosity of a group and $V_{max}$ is the total comoving volume in which a group with $L_{X,i}$ could have been detected above the flux limits of the survey. The total comoving volume is defined by

\begin{equation}
V_{max}(L_X) = \int^{z_{max}}_{z_{min}}\Omega(f_X(L_X,z))\frac{dV(z)}{dz}dz.
\end{equation}

\noindent $\Omega(f_X)$ is the survey solid angle written as a function of the X-ray flux and $dV(z)/dz$ is the differential comoving volume element per steradian. To fit the group XLF, we adopt the Schechter function \citep{1976Schechter},

\begin{equation}
\phi (L_X) = AL_{X}^{-\alpha}{ \rm exp}\left(\frac{-L_X}{L_X^*}\right).
\end{equation}

\noindent This is the parametric representation of the group/cluster XLF, where $L_X^*$ is the characteristic luminosity marking the transition between the power-law and exponential regimes, $\alpha$ is the faint-end slope, $A = \phi^*/L_X^{1-\alpha}$ and $\phi^*$ is the normalization in units of Mpc$^{-3}$.  

\par Typical values of $L_X^*$ are between $10^{44}-10^{45}$ ergs s$^{-1}$, depending on the cosmological model adopted. Since the Bo\"otes sample goes only to a maximum of $L_X=10^{44}$ ergs s$^{-1}$, we add the 160SD low redshift sample from \citet{2004Mullis} to our data to be able to fit $L_X^*$.  In Figure \ref{XLF_fig}, we show the XLF fit (black line) for the X-Bo\"otes (red circles) and 160SD (blue circles) samples. The dashed lines indicate the 1$\sigma$ region of the Schechter fit assuming the errors for $L_X^*$ and $\alpha$ are correlated. The data are plotted at the center of each luminosity bin. The uncertainties on the data points are $\pm 1\sigma$, based on Poissonian errors for the number of groups per luminosity bin \citep{1986Gehrels}.  

\par We find $L_X^*=(1.44\pm1.12)\times 10^{45}$ ergs s$^{-1}$, $\alpha=1.28\pm0.09$ and $\phi^*=(1.65\pm0.32)\times 10^{-6}$ Mpc$^{-3}$ (10$^{44}$ ergs s$^{-1}$)$^{-1}$. The value found for $L_X^*$ is compatible with what was found by \citet{2002Bohringer} for the REFLEX sample ($L_X^* = 1.02^{+0.11}_{-0.09}\times 10^{45}$ ergs s$^{-1}$)\footnote{This value is converted for $\Omega_m = 0.3, \Omega_\Lambda = 0.7$ and $H_0 = 70$ km s$^{-1}$ Mpc$^{-1}$.}.  The normalization $\phi^* = 5.75 \times 10^{-7}$ and the slope $\alpha = 1.69\pm 0.045$ for the REFLEX differ from what we found here, because the REFLEX sample does not extend as low as $L_X = 10^{41}-10^{42}$ ergs s$^{-1}$, the faint end of our XLF.

\par With the XLF parameterized, we can estimate the expected number of groups ($N_{exp}$) for a given range of luminosity inside a given volume. Assuming there is no evolution in the XLF, $N_{exp}$ can be computed by integrating the luminosity function $\phi(L_X)$ over a given volume and a luminosity interval, according to this equation:

\begin{equation}
N_{exp} = \int^{L_{X}^{max}}_{L_{X}^{min}}\int^{z_{max}}_{z_{min}}\phi(L_X)\Omega(f_X(L_X,z))\frac{dV(z)}{dz}dzdL_X.
\end{equation}

\noindent Assuming that the eROSITA all sky survey will be complete to $f_X = 3\times 10^{-14}$ ergs s$^{-1}$ cm$^{-2}$, we adopt $\Omega(f_X(L_X,z)) = 4\pi$. To integrate $N_{exp}$, we consider four intervals of redshift ($0 < z \leq 0.046$, $0.046 < z \leq 0.114$, $0.114 < z \leq 0.238$ and $0.238 < z \leq 0.319$) and three intervals of luminosity ($10^{41} \leq L_X \leq 10^{42}$, $10^{42} \leq L_X \leq 5\times10^{42}$ and $5\times10^{42} \leq L_X \leq 10^{43}$ ergs s$^{-1}$ cm$^{-2}$).  The results, listed in Table \ref{tab_nexp}, show that with eROSITA it will be possible to detect approximately 27,400 groups within $z=0.32$.  Approximately 1800 of these groups between $10^{41} \leq L_X \leq 10^{43}$ ergs s$^{-1}$ will be within $z = 0.046$, which corresponds to a luminosity distance ($D_L$) of 200 Mpc.  Thus, since groups trace the large scale filaments, the eROSITA survey by determining the locations of groups will map the large scale structures and filaments throughout the local universe.

%%%%%%%%%%%%%%%%%%%%%%%%%%%%%%%%%%%%%%%%%%%%%%%%%%%%%%%%%%%%%%%%%%%%%%%%
%%%%%%%%%%%%%%%%%%%%%%%%%%%%%%%%%%%%%%%%%%%%%%%%%%%%%%%%%%%%%%%%%%%%%%%%
%%%%%%%%%%%%%%%%%%%%%%%%%%%%%%%%%%%%%%%%%%%%%%%%%%%%%%%%%%%%%%%%%%%%%%%%
%%%                                                                  %%%
%%%                         CONCLUSIONS                              %%%
%%%                                                                  %%%
%%%%%%%%%%%%%%%%%%%%%%%%%%%%%%%%%%%%%%%%%%%%%%%%%%%%%%%%%%%%%%%%%%%%%%%%
%%%%%%%%%%%%%%%%%%%%%%%%%%%%%%%%%%%%%%%%%%%%%%%%%%%%%%%%%%%%%%%%%%%%%%%%
%%%%%%%%%%%%%%%%%%%%%%%%%%%%%%%%%%%%%%%%%%%%%%%%%%%%%%%%%%%%%%%%%%%%%%%% 

\section{Conclusions}

\par We have used the extended source catalog from \citet{2005Kenter} to build the X-Bo\"otes galaxy group catalog.  Group redshifts are measured from the AGES \citep{2012Kochanek} spectroscopic data. Our final sample comprises 32 systems at $z < 1.75$, with 14 below $z = 0.35$. For groups with at least five galaxy members inside $R_{500}$, we apply a virial analysis to estimate velocity dispersions and to obtain the radii ($R_{200}$ and $R_{500}$) and total masses ($M_{200}$ and $M_{500}$). To derive the group richness and the optical luminosity, we use photometric data from the NDWFS \citep{1999Jannuzi}. The X-ray luminosity for each group is determined from the Chandra X-ray observations. 

\par After measuring the group properties, we examine their performance as proxies for the group mass. It is important to understand how well these observables measure the total mass, because they will often be used as mass proxies in large surveys (e.g. eROSITA). Exploring the scaling relations built with the X-Bo\"otes sample and comparing them with samples from the literature, we find a break in the $L_X-M_{500}$ relation at approximately $M_{500} = 5\times10^{13}$ M$_{\odot}$. A possible explanation for this break can be the dynamical friction, tidal interactions and projection effect which reduce the velocity dispersion values of the galaxy groups.  We also examine the $L_X-\sigma_{gr}$, $L_{opt}-M_{500}$ and $L_{opt}-\sigma_{gr}$ relations. But, due to the large scatter in these relations, it is difficult to determine if there is a break between cluster and group samples.   

\par By extending the Log\textit{N}-Log\textit{S} and the luminosity function to the group regime, we predict the number of groups that the new X-ray survey, eROSITA, will detect per interval of luminosity and distance. eROSITA will observe a total of 27,400 galaxy groups within $z=0.32$ with $L_X$ between 10$^{41}$-10$^{43}$ ergs s$^{-1}$. The galaxy groups will be a powerful tool for eROSITA to map the large scale structure in the local universe.

\acknowledgements

The work was supported in part by Chandra grant, NASA grant, Smithsonian Institute grant and
CsF/CNPq grant 237321/2012-2.

%%%%%%%%%%%%%%%%%%%%%%%%%%%%%%%%%%%%%%%%%%%%%%%%%%%%%%%%%%%%%%%%%%%%%%%%
%%%%%%%%%%%%%%%%%%%%%%%%%%%%%%%%%%%%%%%%%%%%%%%%%%%%%%%%%%%%%%%%%%%%%%%%
%%%%%%%%%%%%%%%%%%%%%%%%%%%%%%%%%%%%%%%%%%%%%%%%%%%%%%%%%%%%%%%%%%%%%%%%
%%%                                                                  %%%
%%%                         BIBLIOGRAPHY                             %%%
%%%                                                                  %%%
%%%%%%%%%%%%%%%%%%%%%%%%%%%%%%%%%%%%%%%%%%%%%%%%%%%%%%%%%%%%%%%%%%%%%%%%
%%%%%%%%%%%%%%%%%%%%%%%%%%%%%%%%%%%%%%%%%%%%%%%%%%%%%%%%%%%%%%%%%%%%%%%%
%%%%%%%%%%%%%%%%%%%%%%%%%%%%%%%%%%%%%%%%%%%%%%%%%%%%%%%%%%%%%%%%%%%%%%%% 

\bibliographystyle{apj}

% Journal name macros

%% \newcommand\mnras{MNRAS}
%% \newcommand\apj{ApJ}
%% \newcommand\apjl{ApJ}
%% \newcommand\apjs{ApJS}
%% \newcommand\nat{Nature}
%% \newcommand\aap{A\&A}
%% \newcommand\aapr{A\&A}
%% \newcommand\araa{ARAA}
%% \newcommand\physrep{PhysRep}
%% \newcommand\aj{AJ}
%% \newcommand\sovast{SovAst}

%%\newcommand\jcap{J.\ Cosmol.\ Astropart.\ Phys.}

%\newcommand\jcap{JCAP}
\newcommand\rvmp{RvMP}

\bibliography{bootes_draft}

\appendix

\section{Extended Sources}

\par In this appendix, we present the catalog of the 14 galaxy groups to $z=0.35$ with at least 5 member galaxies inside $R_{500}$.  For each group, we show the density profile of the AGES galaxies centered in the central coordinates of the galaxy group and the spatial distribution of galaxies in the $\Delta$R.A. vs. $\Delta$Dec diagram.

\begin{figure*}[!htb]
\includegraphics[width=184mm]{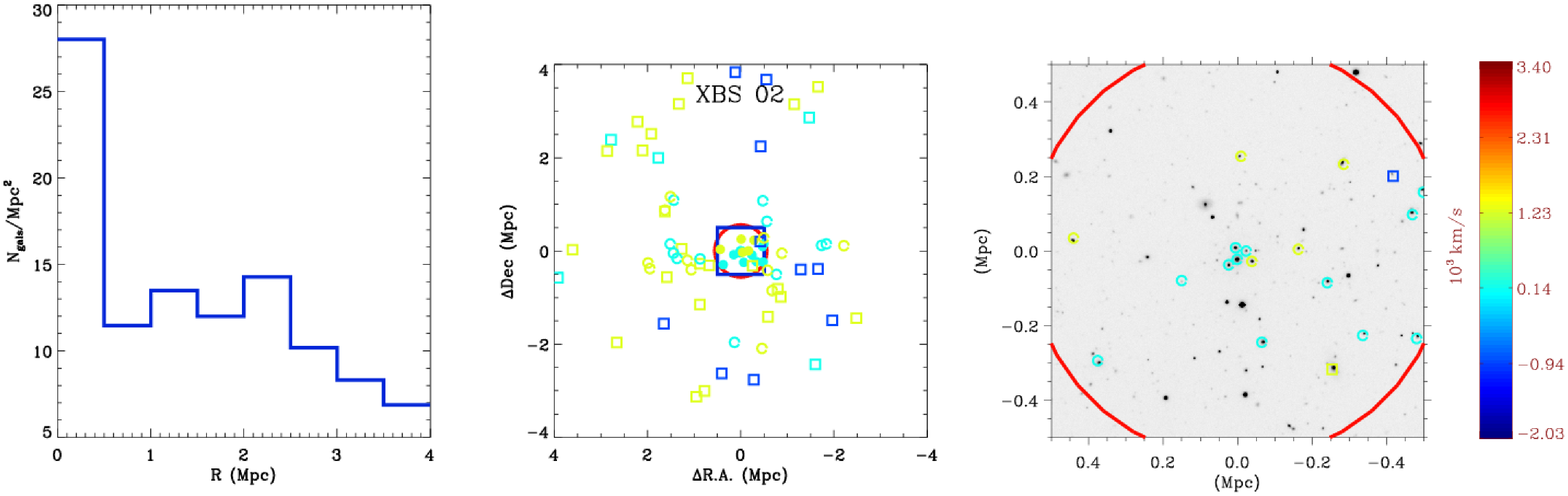}
\includegraphics[width=184mm]{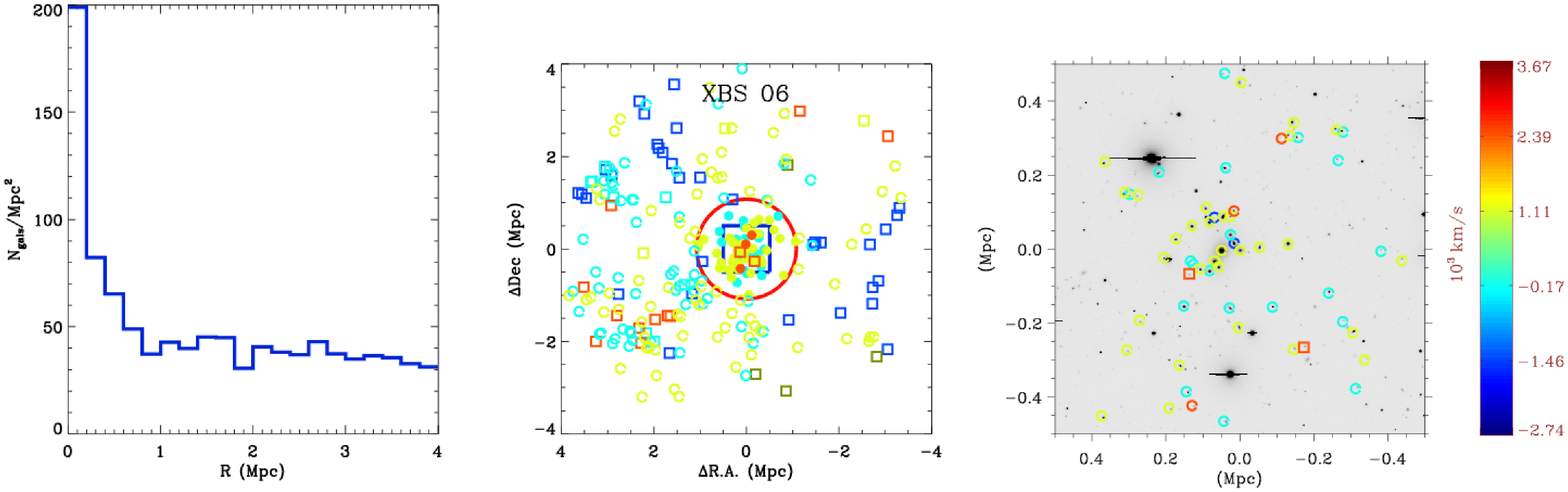}
\caption{The left panel in each row is the density profile of the AGES galaxies centered in the central coordinates of the galaxy group.  The middle panel is the $\Delta$R.A. vs. $\Delta$Dec diagram.  The filled circles are the galaxy members inside $R_{500}$ (big red circle) used to estimate $\sigma$.  The open circles are galaxy members outside $R_{500}$.  The open squares are the rejected interlopers.  The colors of the symbols represent the difference in velocity between the group and the galaxies.  The color scale is represented by the color bar (in units of 10$^3$ km s$^{-1}$) on the right side of the the right panel.  The right panel is the zoom in region of the 1 Mpc blue box of the middle panel with I band NDWFS image in the background.  The open circles are the galaxy members inside $R_{500}$ used to estimate $\sigma$.  The open squares are the rejected interlopers.  The symbol colors follow the same scale of the color bar on the right.}
\label{finding_charts}
\end{figure*}

\begin{figure*}
\begin{center}
\includegraphics[width=184mm]{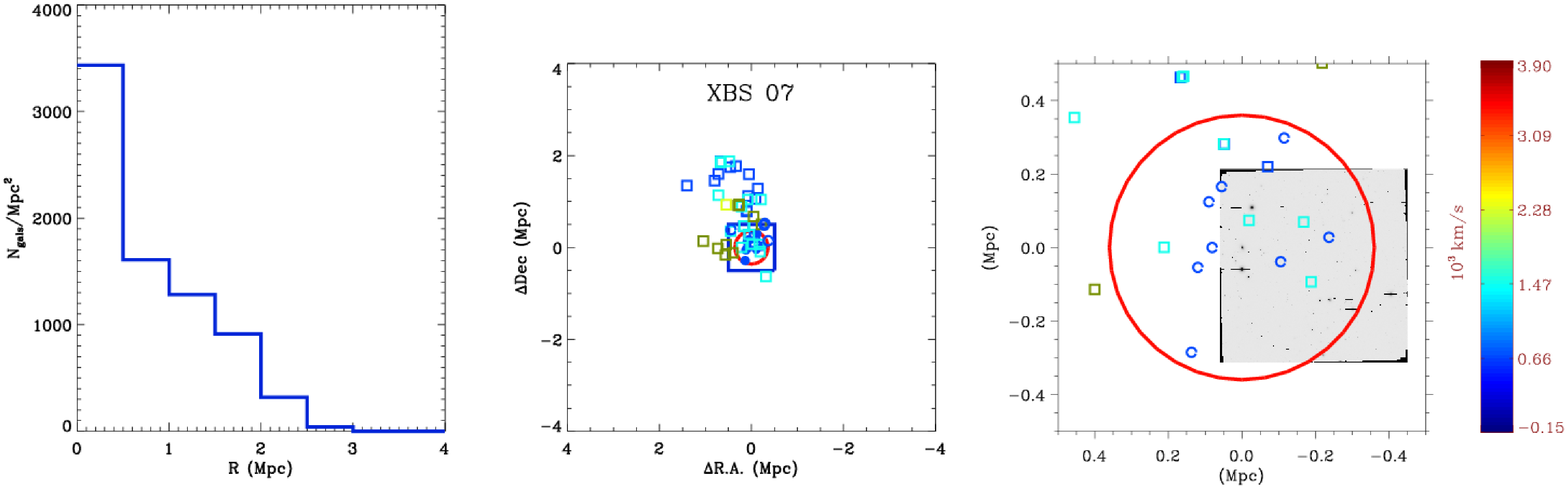}
\includegraphics[width=184mm]{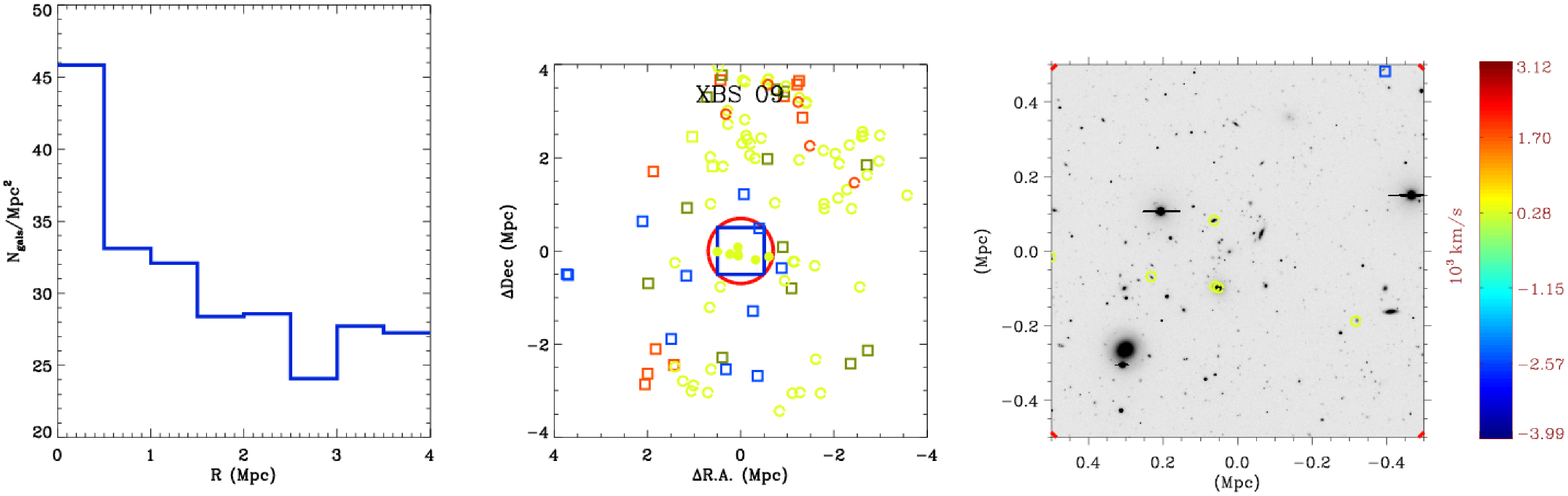}
\includegraphics[width=184mm]{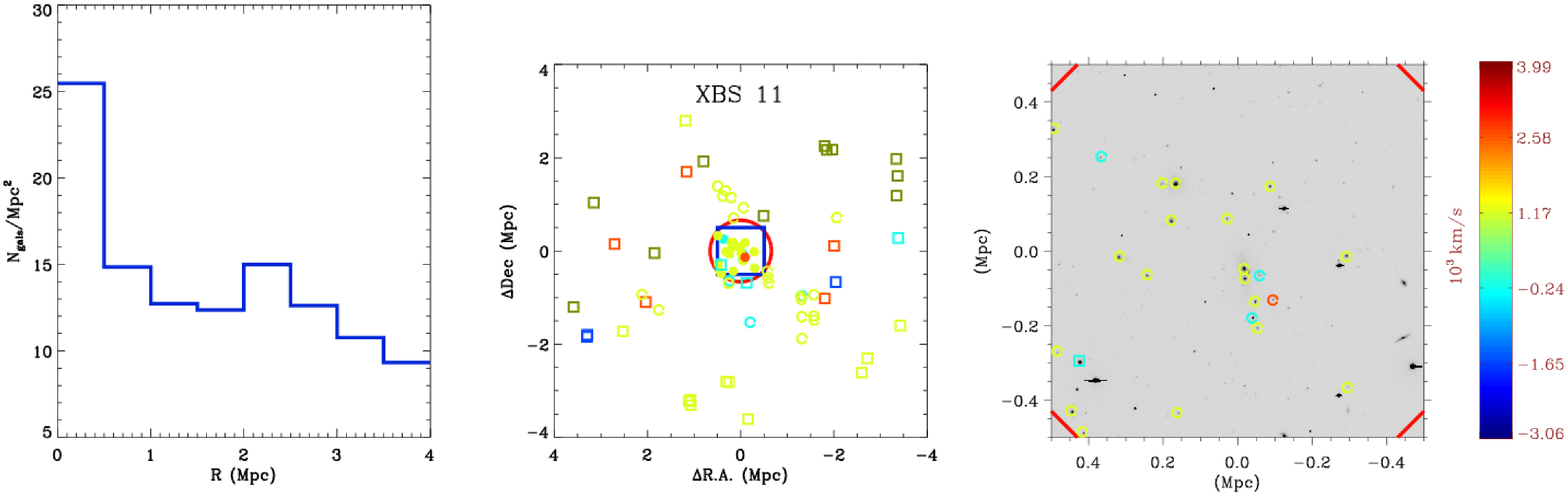}
\end{center}
\caption{Continuation of Figure \ref{finding_charts}.}
%\label{}
\end{figure*}

\begin{figure*}
\begin{center}
\includegraphics[width=184mm]{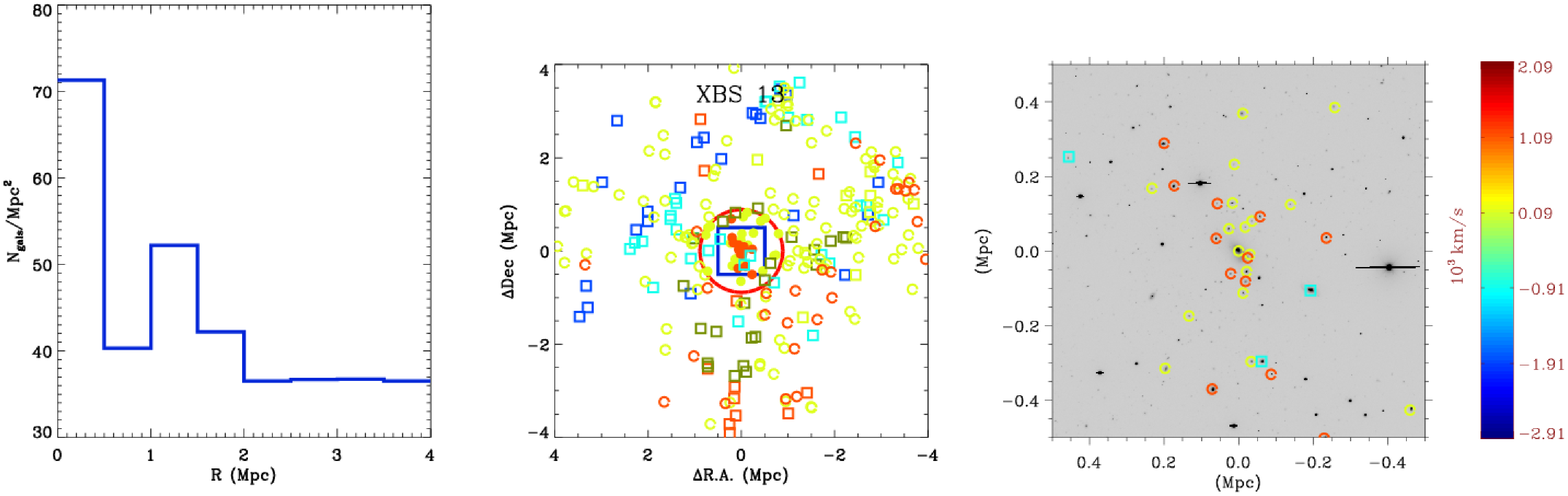}
\includegraphics[width=184mm]{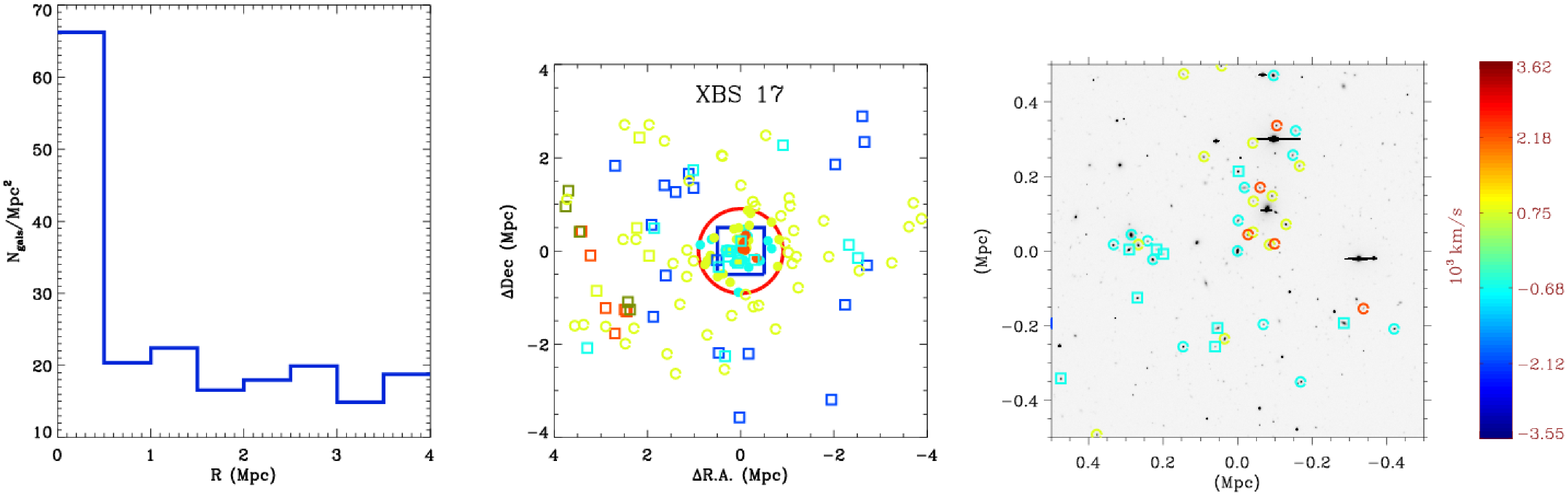}
\includegraphics[width=184mm]{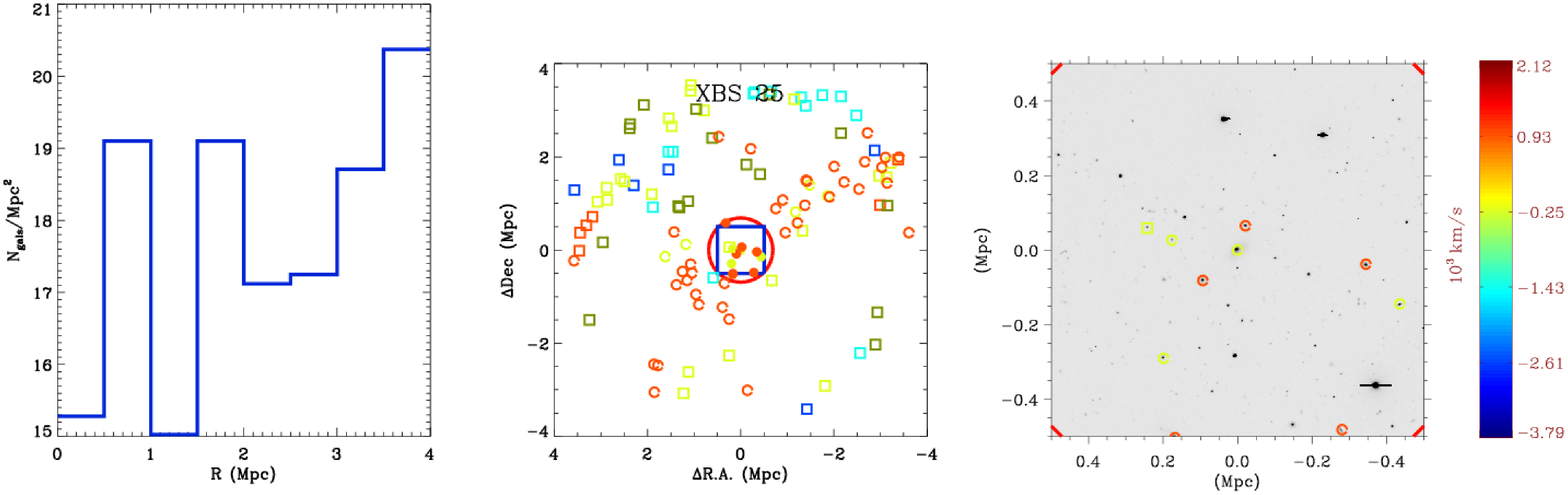}
\end{center}
\caption{Continuation of Figure \ref{finding_charts}}
%\label{}
\end{figure*}

\begin{figure*}
\begin{center}
\includegraphics[width=184mm]{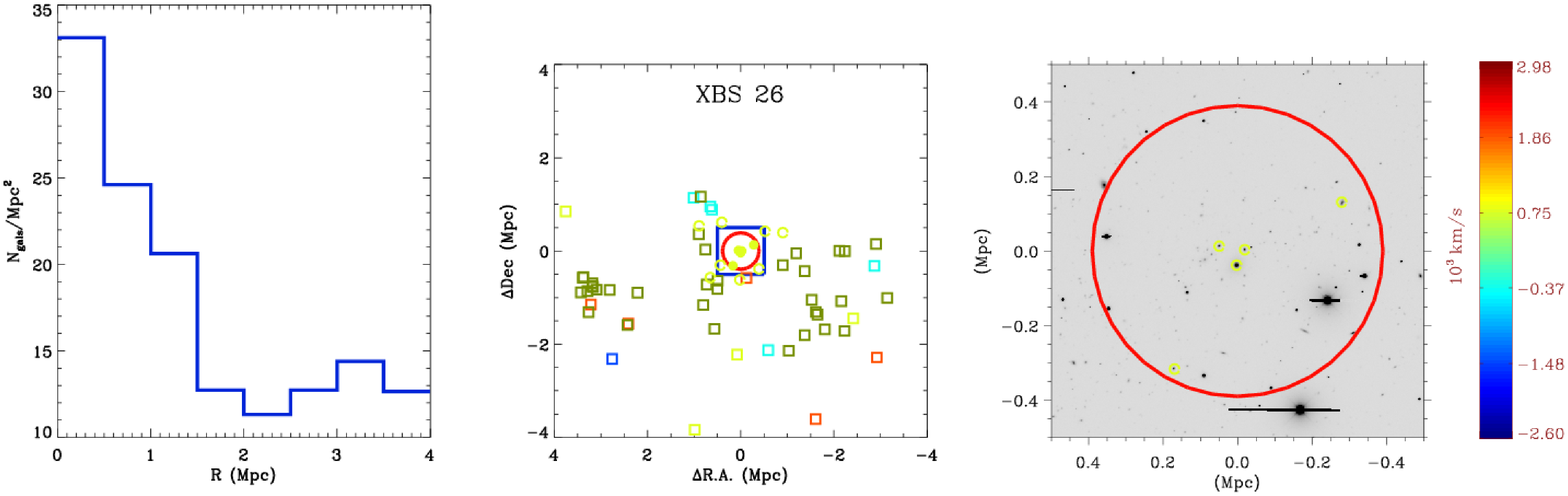}
\includegraphics[width=184mm]{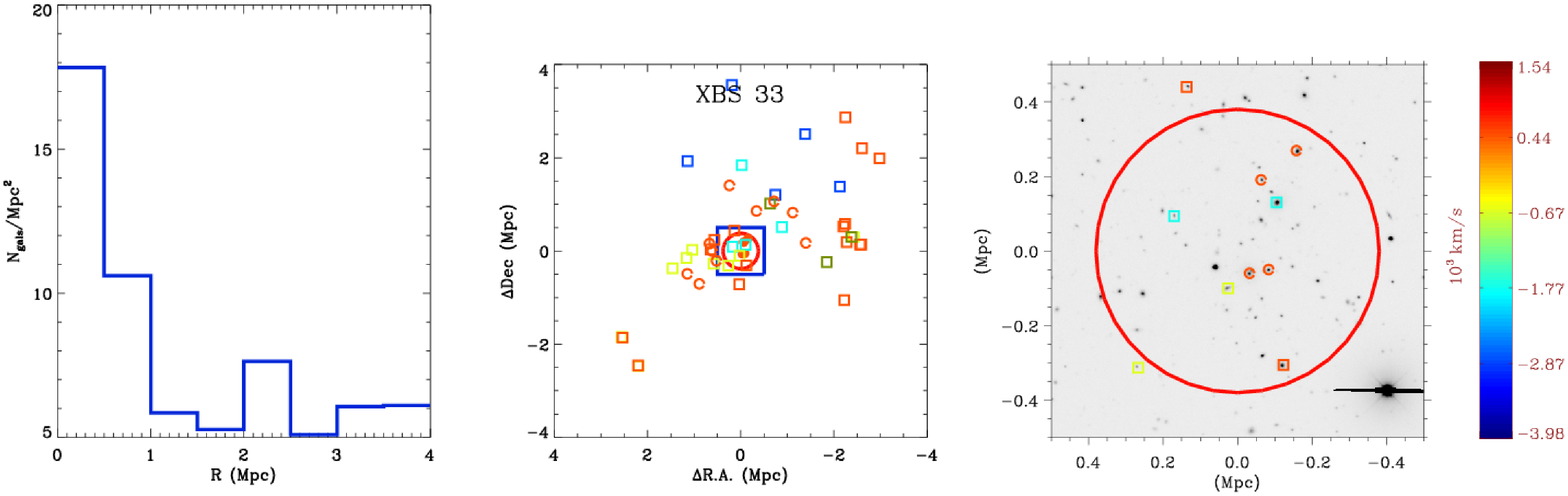}
\includegraphics[width=184mm]{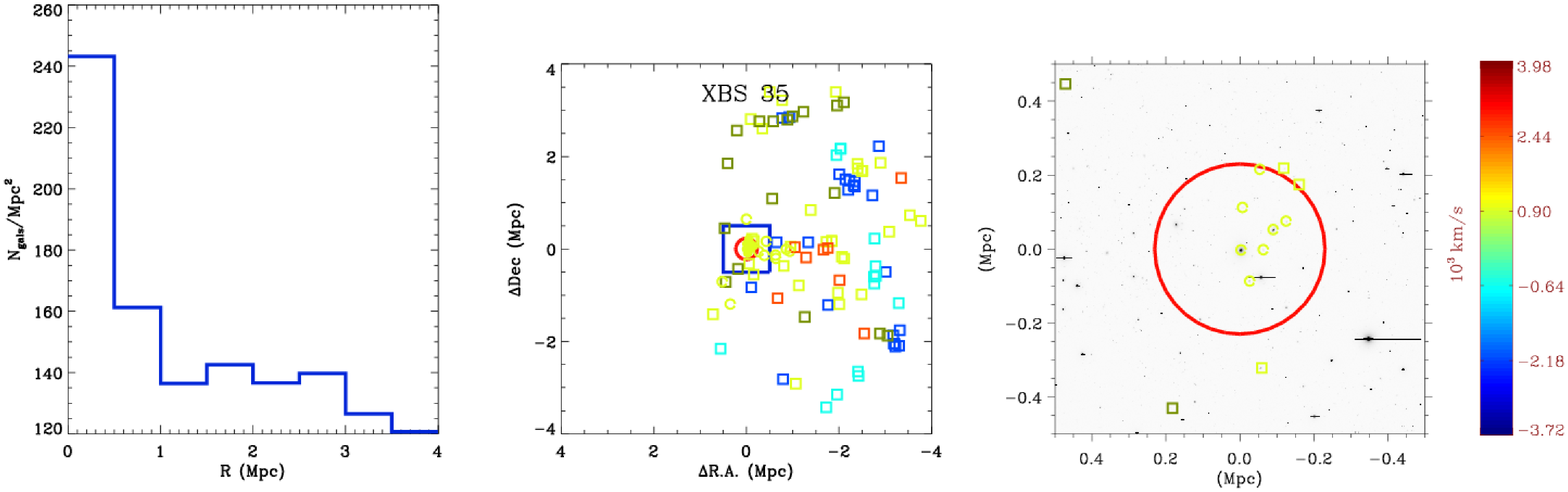}
\end{center}
\caption{Continuation of Figure \ref{finding_charts}}
%\label{}
\end{figure*}

\begin{figure*}
\begin{center}
\includegraphics[width=184mm]{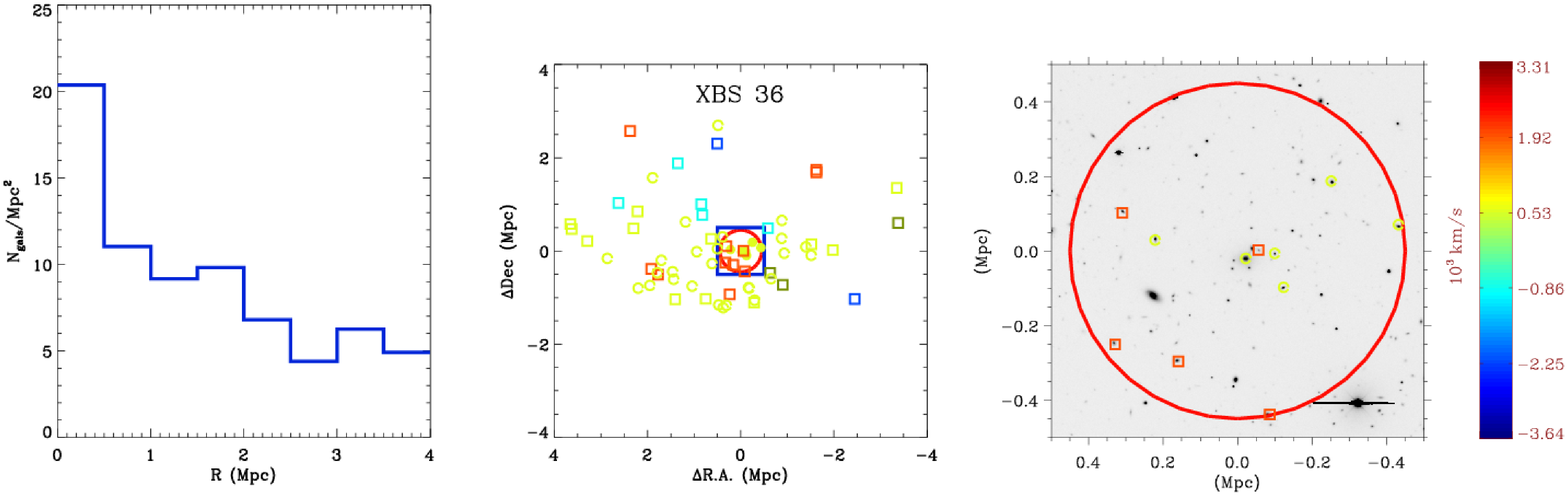}
\includegraphics[width=184mm]{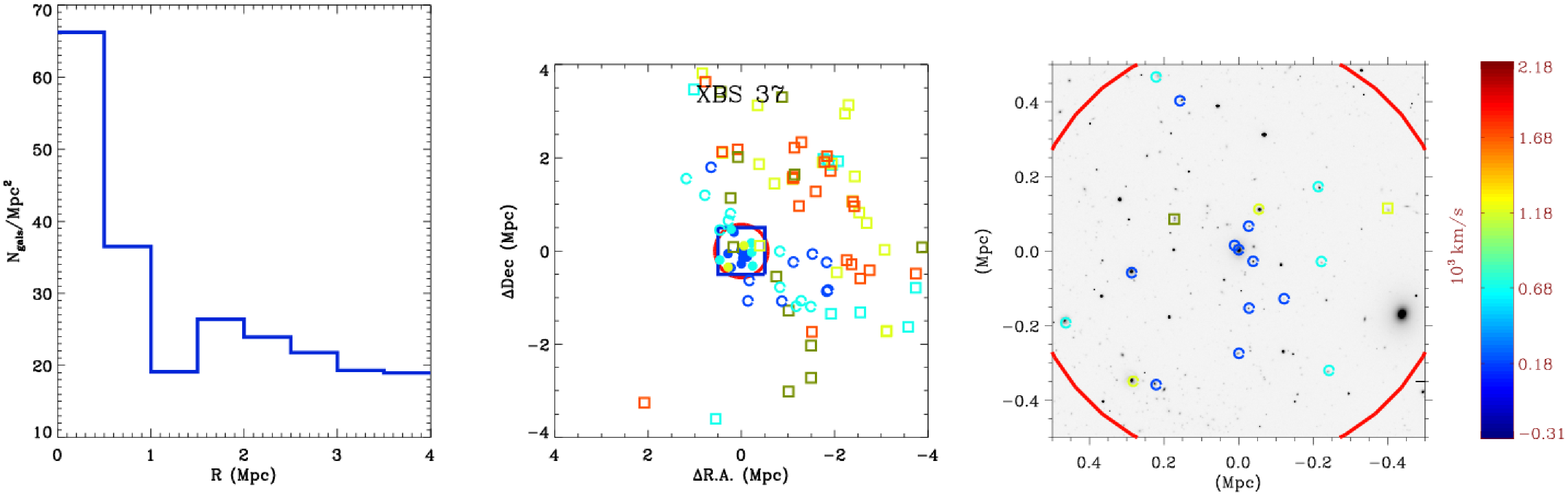}
\includegraphics[width=184mm]{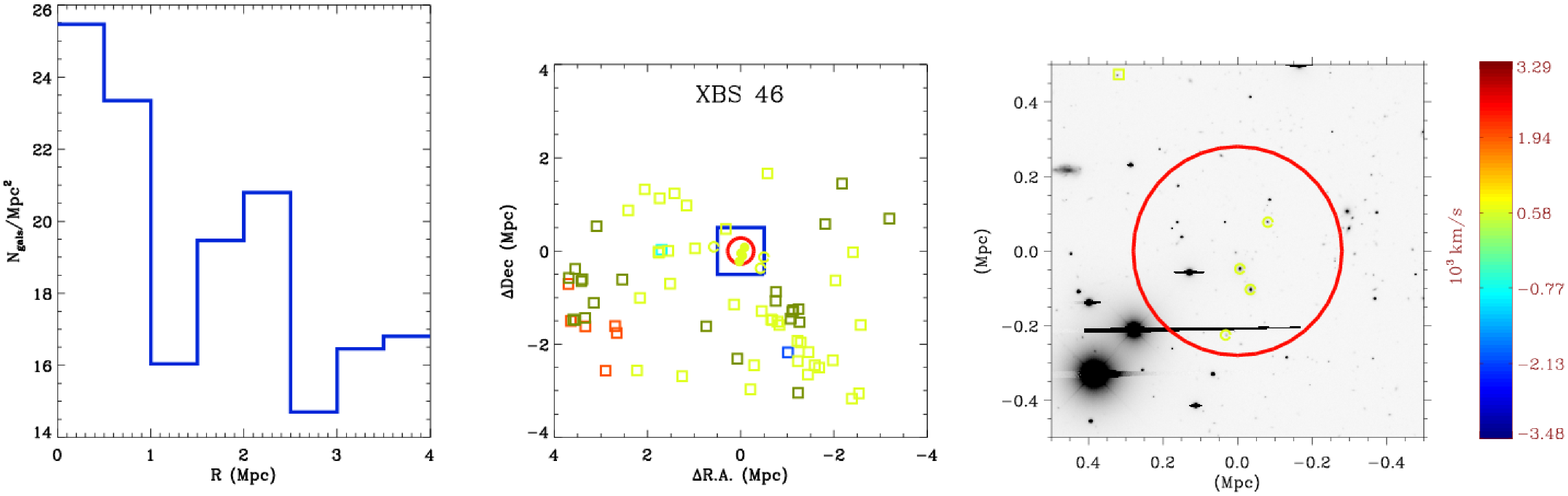}
\end{center}
\caption{Continuation of Figure \ref{finding_charts}}
%\label{}
\end{figure*}

\end{document}